\documentclass[aps,pre,twocolumn,superscriptaddress,showpacs]{revtex4}
\usepackage{amsmath,amssymb,amsfonts}
\usepackage{bbm}
\usepackage{graphicx}

\newcommand\ImgI{\mathrm{i}}
        
\newcommand\Exp{\mathrm{e}}

\newcommand\RE{\mathrm{Re}\,}
\begin{document}

\title{Universal spectral statistics in Wigner-Dyson, chiral and Andreev
  star graphs II: semiclassical approach}

\author{Sven Gnutzmann} 
\email{sven@gnutzmann.de}
\affiliation{Institut f\"ur Theoretische Physik, Freie Universit\"at
  Berlin, Arnimallee 14, 14195 Berlin, Germany} 
\author{Burkhard Seif}
\email{bseif@thp.uni-koeln.de} 
\affiliation{Institut f\"ur
  Theoretische Physik, Universit\"at zu K\"oln, Z\"ulpicher Str.\ 77,
  50937 K\"oln}
 
\begin{abstract} 
  
  A semiclassical approach to the universal ergodic spectral
  statistics in quantum star graphs is
  presented for all known ten symmetry classes of quantum systems.
  The approach is based on periodic orbit theory, the exact
  semiclassical trace formula for
  star graphs and on diagrammatic techniques. The
  appropriate spectral form factors are calculated
  upto one order beyond the diagonal and self-dual approximations.
  The results are in accordance with the
  corresponding random-matrix theories which supports
  a properly generalized Bohigas-Giannoni-Schmit conjecture.

\end{abstract}      


\maketitle

\section{Introduction}

Since Bohigas, Giannoni and Schmit conjectured in their
in a seminal paper \cite{BGS} that the spectral
fluctuations in quantum systems with a chaotic classical
counterpart follow the predictions of the Gaussian
random-matrix ensembles GUE, GOE, or GSE (depending
on the behavior of the system under time-reversal and spin rotations)
a lot of numerical data has been gathered that strongly support
this conjecture (see \cite{Guhr,bible} and references therein).
While an analytic proof of the conjecture
and a precise statement of its limits
is still lacking there has been a continuous advance
in understanding the
universality of spectral statistics.
The main tool in the semiclassical approach is Gutzwiller's
trace formula \cite{Gutzwiller} 
which expresses the fluctuating part of
the density of states as a sum over classical periodic orbits.
The main object of interest is the spectral form factor which
is the Fourier transform of the spectral two-point
correlation function (below we will call this the second-order
form factor) which is expressed semiclassically
via the trace formula as a sum over pairs of periodic
orbits which share the same action.
In the diagonal approximation introduced by Berry \cite{diagonal}
only those pairs for which both periodic orbits
are either equal or the time-reverse of the other are
summed over. The assumption
of hyperbolic chaos is then sufficient to prove that the
leading linear order in a short-time expansion for
a quantum system follows the random-matrix predictions.
Recently Sieber and Richter \cite{Sieber} started new progress for a
semiclassical approach beyond the diagonal approximation 
with the observation that self-crossing
trajectories in a billiard of constant negative curvature have
a partner orbit of the same action that avoids this self-crossing.
In the form factor these Sieber-Richter pairs give the
quadratic order in time as predicted by the GOE. 
Their approach has been generalized to general hyperbolic billiards
\cite{Sebastian} and later in a phase space approach 
to hyperbolically chaotic Hamiltonian systems with two
\cite{Dominique,Marko} and finally any number of
degrees of freedom \cite{SebDomMark}. Finally, the cubic
order in the short-time expansion has been calculated 
\cite{tauhochdrei}.

The first non-vanishing order in $\tau$ is known as the 
\emph{diagonal approximation} in the Wigner-Dyson classes
which is linear in $\tau$.
For the chiral and Andreev classes the first non-vanishing order is
$\tau^0$ and has
been called \emph{the self-dual approximation} \cite{us}.
So far the fidelity to the predictions
of Gaussian random-matrix theory has been derived
for the classes $C$ and $C$I -- here, we will give a complete account
of for star graphs in all symmetry classes. Though we
restrict to graphs here, the following calculation show what
types of periodic orbits and which of their properties are
responsible for universality in more general Hamiltonian system.  

The next to leading order (the \emph{weak localization corrections})
will also be calculated for all ten ensembles of star graphs.
For fully connected Wigner-Dyson graphs in class $A$I (GOE) these have 
recently been
calculated to order $\tau^3$ (based on much earlier work on $\tau^2$)
by Berkolaiko \textit{et al} \cite{Greg}. 
These authors generalized the Sieber-Richter approach to graphs and 
introduced diagrammatic techniques similar to those used in this paper.
The relation between the corresponding expansion for quantum systems 
in class $A$II (GSE) and the $A$I (GOE) expansion has been considered receltly
\cite{heuslerGSE,harrisonGSE}.

The corresponding ensembles of star graphs
have been constructed in 
the first paper of this series \cite{usI}. Each of these
ensembles
obeys the symmetry conditions of one class
in the ten-fold classification of quantum systems. 
There, we have also introduced the first-order and second-order 
spectral form factors
as the Fourier transforms of the fluctuating part of
the density of states and the two-point correlation function
and related them to them to the scattering matrix if
the star graphs via an exact semiclassical trace formula.
We have also shown numerically that these spectral
form factors follow the 
predictions of the corresponding Gaussian random-matrix ensembles.
The second-order form factor of graphs in the Wigner-Dyson classes
follow the predictions of the well-known Wigner-Dyson ensembles
GUE, GOE, GSE. For the remaining seven ``novel'' symmetry
classes
the first-order form factor for star graphs coincides 
with the corresponding Gaussian random-matrix prediction.
For the novel ensembles the fidelity to Gaussian random-matrix
prediction is the content of a properly generalized
Bohigas-Giannoni-Schmit conjecture \cite{usI}.

The (generalized) Bohigas-Giannoni-Schmit conjecture
states that the spectral fluctuations
of a \emph{single} classically chaotic physical system 
follow the predictions of Gaussian random-matrix theories. 
Here, we have explicitly introduced \emph{ensembles}
of star graphs.
It has been shown however \cite{Barra} that ensemble
averages over certain 
phases in graphs are equivalent to a spectral average
of a \emph{single} spectrum in a graph with incommensurate bond
lengths. 

For the first-order form factor
for chiral and Andreev star graphs the generalized
Bohigas-Giannoni-Schmit conjecture is slightly weaker in as much
as not a single spectrum is conjectured to follow the random-matrix
predictions but a one-parameter average over different
values of an effective $\hbar$ (an average over different values
of the Fermi level or different quantizations
of the same system). We leave it open here to what extend the
ensemble averages for the graphs in the novel symmetry classes
are equivalent to such a one-parameter average.

In section \ref{sec:diagrams} we introduce the diagrammatic
representation of the form factors: the section starts with
a general description in \ref{sec:general_diagrams}, there we give the
vertex (``d-vertex'') and bond (``line'')
contributions to a diagram for each of the ten symmetry classes in
\ref{sec:symmetries} and also give the
diagrammatic expansion of the ensemble averaged 
form factors in \ref{sec:expansion}.
In section \ref{sec:short-time} we calculate the diagonal 
and self-dual approximations
and one order beyond for the form factors of the ten ensembles:
this section first introduces a systematic diagrammatic short-time
expansion of the form factors in \ref{sec:WD_short_time} and then explicitly
gives all diagrams of the leading and next-to-leading order
which are calculated explicitly in \ref{sec:diagonal_selfdual}.

\section{The diagrammatic representation of form factors for star graphs}
\label{sec:diagrams}

Star graphs are simple quantum systems with an exact semiclassical
trace formula for the density of states \cite{Kottos}. They consist of
$V$ vertices connected by $B$ bonds of length $L_i$. 
A particle propagates freely
on the bond and is scattered at the vertices according to prescribed
unitary vertex scattering matrices. 

In a star graph $B$ bonds emanate from one central vertex and connect
it to $B$ peripheral vertices \cite{usI,Kottos,gregonstars}. 
We have generalized
previous star graph models by allowing for a wave function with $\mu$
components. In our model all bonds have the same length
and the free propagation along the bonds and the
scattering at the central vertex do not mix the components (but 
have to obey some symmetry conditions). The central vertex 
scattering is thus a unitary $\mu B\times \mu B$ matrix of
the form $\mathcal{S}_{C,\alpha j,\alpha',j'}=\delta_{\alpha \alpha'}
\mathcal{S}_{C,jj'}^{(\alpha)}$ \cite{usI}. Here, $j,j'=1,\dots,B$
is an index for the bonds and $\alpha,\alpha'=1,\dots,\mu$ counts
the wave function components.   
In addition, propagation along the bonds
and the central scattering are time-reversal invariant. The proper
choice of the scattering process at the peripheral vertices fixes
the symmetry class (and breaks time-reversal if necessary).
All peripheral scattering processes can be described by a single
unitary $\mu B\times \mu B$ matrix of
the form $\mathcal{S}_{P,\alpha j,\alpha',j'}=\delta_{j j'}
\sigma_{\alpha \alpha'}^{(j)}$ \cite{usI}.
For graphs in the novel symmetry classes this process involves
an equivalent of
\emph{Andreev scattering} (electron-hole conversion).
 
\subsection{\label{sec:general_diagrams} Diagrammatic representation
  of the form factors}

Prescribing the central and peripheral scattering vertex matrices
$\mathcal{S}_C$ and $\mathcal{S}_P$ leads to a quantization of the graph.
Their product is the reduced bond scattering matrix 
$\mathcal{S}_B\equiv\mathcal{S}_P\mathcal{S}_C$.
The density of states is represented exactly by the semiclassical
trace formula
\begin{equation}
  d(\kappa) = 1+ \frac{2}{\mu B}\, \RE \sum_{n=1}^{\infty}
  \Exp^{\ImgI 2 \pi \kappa \frac{ g n}{\mu B}} s_n.
  \label{eq:trace}
\end{equation}
Here, $\kappa$ is the wave number in units of the (macroscopic)
mean level spacing and
\begin{equation}
  s_n=\mathrm{tr}\, \tilde{\mathcal{S}}_B^n
  =\sum_{j_i,\alpha_i} \sigma_{\alpha_1 \alpha_n}^{(j_1)}
  \mathcal{S}_{C,j_1 j_n}^{(\alpha_n)} \dots
  \sigma_{\alpha_2 \alpha_1}^{(j_2)} \mathcal{S}_{C,j_2
    j_1}^{(\alpha_1)}
  \label{eq:sn}
\end{equation}
is the trace of the $n$-th power of the reduced bond scattering matrix. 
The latter may be represented by diagrams 
\cite{Greg}
\begin{equation}
  s_n= 
  \begin{array}{c}
    \includegraphics[scale=0.8]{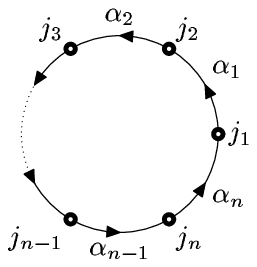}
  \end{array}.
  \label{eq:trace_sn_diag}
\end{equation}
In order to distinguish  
between a bond or vertex in a star graph
and in a diagram we will use the
terms \emph{lines} and \emph{d-vertices} for the diagrams and reserve
\emph{bonds} and \emph{vertices} for star graphs.  
The d-vertices in the above diagram correspond to the
peripheral vertices in
the star graph which are visited one after another.
Each d-vertex contributes
a factor
\begin{equation}
  \begin{array}{c}
    \includegraphics[scale=0.8]{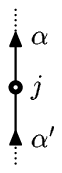}
  \end{array}
  =\sigma^{(j)}_{\alpha \alpha'}\equiv \mathcal{S}_{P,\,j \alpha,j' \alpha'}
\end{equation}
to the diagram. 
This describes the scattering of a particle that moves outwards on
bond $j$ in the state $\alpha'$ to a particle moving inwards on the
same bond in state $\alpha$.  The lines in the diagram carry a factor
\begin{equation}
  \begin{array}{c}
    \includegraphics[scale=0.8]{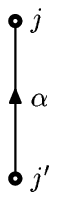}
  \end{array}
  =\mathcal{S}_{C,j j'}^{(\alpha)}\equiv
  \mathcal{S}_{C,\, j \alpha, j' \alpha}.
\end{equation}
A diagram is calculated by summing over all indices
all indices $j_k=1,\dots,B$ and
$\alpha_k=1,\dots,\mu$ -- obviously for
\eqref{eq:trace_sn_diag} one arrives back at \eqref{eq:sn}.
We have defined the first-order form factor \cite{usI}
as the Fourier transform of the (ensemble averaged)
fluctuating part of the density of states. The first-order form
factor in discrete time $\tau=\frac{g n}{\mu B}$ ($g=2$ in
classes $A$II, $D$III, and $C$II which have
Kramers' degeneracy, else $g=1$) is upto a constant given by
the ensemble average 
\begin{equation}
  K_{1,n}= \frac{2}{g}\, \left\langle s_n \right\rangle.
\end{equation} 
After an additional time average (over a small interval $\Delta n\ll B$) 
which is needed for
comparison with random-matrix theory the first-order form factor 
(in continuous time) is
$ K_1(\tau=\frac{gn}{\mu B})=  
\frac{1}{\Delta n}\sum_{n'=n}^{n+\Delta n-1} K_{1,n}$.
Equivalently there is a discrete and continuous time second-order 
form factors. In the Wigner-Dyson case they are given by
\begin{equation}
  \begin{array}{rl}
    K_{2,n}&=\frac{1}{B}\,\langle s_n s_n^*\rangle \\
    &= \frac{1}{B}\, \left\langle
      \begin{array}{c}
        \includegraphics[scale=0.8]{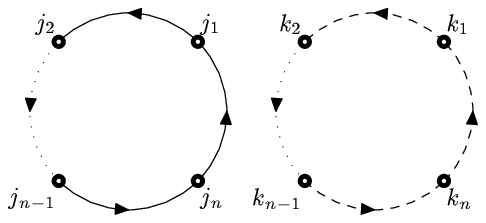}
      \end{array}
    \right\rangle
  \end{array}
  \label{eq:AIff}
\end{equation}
and
$K_2(\tau=\frac{gn}{\mu  B})=\frac{1}{\Delta n}
\sum_{n'=n}^{n+\Delta n-1} K_{2,n}$.
In the diagram for $s_n^*$ we have indicated the complex conjugation by
dashed lines. Most contributions to $s_n$ and $|s_n|^2$ do not
survive the ensemble average. The remaining contributions can be
written as a sum over various diagrams involving the same lines
and d-vertices but have less summation indices.
This expansion will be explained in section \ref{sec:expansion}.

\subsection{\label{sec:symmetries} 
  The line and d-vertex factors for the ten symmetry classes}

In the previous section we have not specified the central and
peripheral scattering matrices for the different symmetry classes.
Let us now give the explicit d-vertex factors and line factors
for each of the ten symmetry classes that are equivalent to the
construction in \cite{usI}.
The lines in the diagram correspond to the central
scattering process. In our model each components of
the wave function is either scattered in the center by the $B \times B$
discrete Fourier transform matrix $\mathcal{S}_{DFT, kl}=
\frac{1}{\sqrt{B}}\Exp^{\ImgI 2 \pi \frac{kl}{B}}$ or by
its complex conjugate for which we will use the lines
\begin{equation}
  \begin{array}{c}
    \includegraphics[scale=0.8]{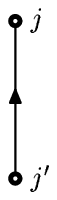}
  \end{array}=\frac{1}{\sqrt{B}}
  \Exp^{2\pi \ImgI \frac{jj'}{B}}
  \qquad\text{and}\qquad
  \begin{array}{c} 
    \includegraphics[scale=0.8]{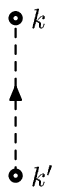}
  \end{array}=
  \frac{1}{\sqrt{B}}
  \Exp^{-2\pi \ImgI \frac{kk'}{B}}. 
  \label{eq:line_contrib}
\end{equation}
For some classes the line carries an additional index for the
component of the wave function -- the line factor however does not depend
on it. 

In the Wigner-Dyson classes only full lines exist
in the representation of $s_n$ while there are only dashed lines in $s_n^*$.
The wave-function has only one component in class $A$I.
It has two components
in class $A$ which we will call ``spin up'' with the
symbol $\uparrow$ and ``spin down'' with the symbol $\downarrow$
for convevenience. Finally, it has a four components in class $A$II --
in addition to the spin labels $\uparrow,\downarrow$ we
use the symbols $\Uparrow,\Downarrow$ and call the latter 
``iso-spin'' up and down for convenience.
The three star graph ensembles in the Wigner-Dyson classes 
are defined by the d-vertex factors to be given now. We only give
the d-vertex factors for in- and outgoing full lines. The corresponding 
factors for d-vertices connected to dashed lines is just the 
complex conjugate. 
In class $A$I each d-vertex carries a random phase factor
\begin{equation}
  \text{$A$I:}\qquad
  \begin{array}{c}
    {\includegraphics[scale=0.8]{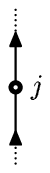}}
  \end{array}=
  \Exp^{\ImgI \beta_j}.
\end{equation}
In class $A$I we have four different scattering processes
corresponding to incoming and outgoing spin directions
\begin{equation}
  \text{$A$:}\qquad
  \begin{array}{rrl}
    \begin{array}{c}
      \includegraphics[scale=0.8]{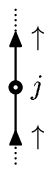} 
    \end{array}=\frac{\Exp^{\ImgI (\beta_j+\gamma_j)}}{\sqrt{2}}&
    \qquad&
    \begin{array}{c}
      \includegraphics[scale=0.8]{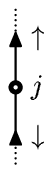} 
    \end{array}=\frac{\Exp^{\ImgI (\beta_j+\delta_j)}}{\sqrt{2}}
    \\
    \begin{array}{c}
      \includegraphics[scale=0.8]{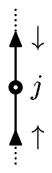} 
    \end{array}=\frac{\Exp^{\ImgI (\beta_j-\delta_j)}}{\sqrt{2}}&
    \qquad &
    \begin{array}{c}
      \includegraphics[scale=0.8]{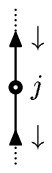} 
    \end{array}=\frac{\Exp^{\ImgI (\beta_j-\gamma_j)}}{\sqrt{2}}.
  \end{array}
\end{equation}
Thus spin either flips with probability $\frac{1}{2}$.
In class $A$II iso-spin always flips at a d-vertex
while spin flips with probability $\frac{1}{2}$. Thus
there are altogether eight different processes
\begin{equation}
  \text{$A$II:}\qquad
  \begin{array}{rrl}
    \begin{array}{c}
      \includegraphics[scale=0.8]{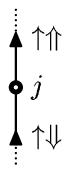} 
    \end{array}=& 
    \begin{array}{c}
      \includegraphics[scale=0.8]{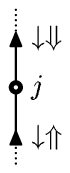} 
    \end{array}&=
    \frac{\Exp^{\ImgI (\beta_j+\gamma_j)}}{\sqrt{2}}\\
    \begin{array}{c}
      \includegraphics[scale=0.8]{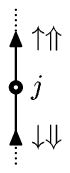} 
    \end{array}=&
    {\scriptstyle(-1)\times}
    \begin{array}{c}
      \includegraphics[scale=0.8]{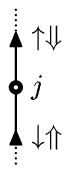} 
    \end{array}&=
    \frac{\Exp^{\ImgI (\beta_j+\delta_j)}}{\sqrt{2}}\\
    \begin{array}{c}
      \includegraphics[scale=0.8]{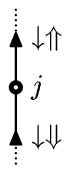} 
    \end{array}=& 
    \begin{array}{c}
      \includegraphics[scale=0.8]{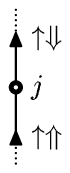} 
    \end{array}&=
    \frac{\Exp^{\ImgI (\beta_j-\gamma_j)}}{\sqrt{2}}\\
    {\scriptstyle (-1)\times}
    \begin{array}{c}
      \includegraphics[scale=0.8]{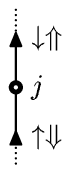} 
    \end{array}
    =&
    \begin{array}{c}
      \includegraphics[scale=0.8]{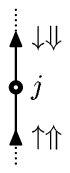} 
    \end{array}
    &= \frac{\Exp^{\ImgI (\beta_j-\delta_j)}}{\sqrt{2}}.
  \end{array}
  \label{eq:vert_AII}
\end{equation}

For the novel symmetry classes the wave function has either two 
(classes $C$ and $C$I)
or four components ($D$, $D$III, $A$III, $BD$I, and $C$II). In either case
the components are divided in ``electron'' and ``hole''  components.
Electrons are represented by full lines and holes by dashed lines.
Additionally for the four-component
wave functions, electrons (holes) have a ``spin'' up and down
component. For each ensemble of star graphs the peripheral
scattering involves complete electron-hole conversion (Andreev scattering).
Thus, each d-vertex is connected to one dashed and one full line.
For the
graphs in the classes $C$ and $C$I they are two different scattering
processes at a d-vertex one for an incoming electron
the other for an incoming hole.
The corresponding d-vector factors are given by
\begin{equation}
  \text{$C$, $C$I:}\qquad
  \begin{array}{c}
    \includegraphics[scale=0.8]{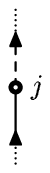}
  \end{array}
  =-\Exp^{- \ImgI \beta_j}\qquad
  \begin{array}{c}
    \includegraphics[scale=0.8]{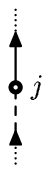}
  \end{array}
  =\Exp^{\ImgI \beta_j}
\end{equation}
where $0\le \beta_j< 2\pi$ is a random phase
in class $C$ and $\beta_j=0$ or
$\beta_j=\pi$ with equal probability in class $C$I. 
In the remaining classes spin flips with probability $\frac{1}{2}$
at each d-vertex. There are altogether eight scattering processes
and their d-vertex
factors are given by
\begin{equation}
  \begin{array}{lccccc}
    &\text{$D$}&\text{$D$III}&\text{$A$III}&\text{$BD$I}&\text{$C$II}\\
    \begin{array}{c}
      \includegraphics[scale=0.7]{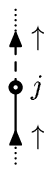} 
    \end{array}
    &\;\;\frac{\Exp^{-\ImgI\beta_j}}{\sqrt{2}}\;\;
    &\;\;\frac{\Exp^{-\ImgI\beta_j}}{\sqrt{2}}\;\;
    &\;\;\frac{\tau_j}{\sqrt{2}}\;\;
    &\;\;\frac{\sigma_j}{\sqrt{2}}\;\;
    &\;\;\frac{-\sigma_j}{\sqrt{2}}\;\;
    \\
    \begin{array}{c}
      \includegraphics[scale=0.7]{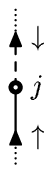} 
    \end{array}
    &\;\;\frac{-\Exp^{\ImgI\gamma_j}}{\sqrt{2}}\;\;
    &\;\;\frac{-\ImgI \sigma_j}{\sqrt{2}}\;\;
    &\;\;\frac{\Exp^{-\ImgI\gamma_j}}{\sqrt{2}}\;\;
    &\;\;\frac{\Exp^{\ImgI\beta_j}}{\sqrt{2}}\;\;
    &\;\;\frac{-\Exp^{-\ImgI\beta_j}}{\sqrt{2}}\;\;
    \\
    \begin{array}{c}
      \includegraphics[scale=0.7]{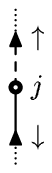} 
    \end{array}
    &\;\;\frac{\Exp^{-\ImgI\gamma_j}}{\sqrt{2}}\;\;
    &\;\;\frac{-\ImgI \sigma_j}{\sqrt{2}}\;\;
    &\;\;\frac{\Exp^{\ImgI\gamma_j}}{\sqrt{2}}\;\;
    &\;\;\frac{\Exp^{-\ImgI\beta_j}}{\sqrt{2}}\;\;
    &\;\;\frac{-\Exp^{\ImgI\beta_j}}{\sqrt{2}}\;\;\\
    \begin{array}{c}
      \includegraphics[scale=0.7]{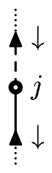} 
    \end{array}
    &\;\;\frac{\Exp^{\ImgI\beta_j}}{\sqrt{2}}\;\;
    &\;\;\frac{\Exp^{\ImgI\beta_j}}{\sqrt{2}}\;\;
    &\;\;\frac{-\tau_j}{\sqrt{2}}\;\;
    &\;\;\frac{-\sigma_j}{\sqrt{2}}\;\;
    &\;\;\frac{\sigma_j}{\sqrt{2}}\;\;\\
    \begin{array}{c}
      \includegraphics[scale=0.7]{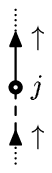} 
    \end{array}
    &\;\;\frac{\Exp^{\ImgI\beta_j}}{\sqrt{2}}\;\;
    &\;\;\frac{\Exp^{\ImgI\beta_j}}{\sqrt{2}}\;\;
    &\;\;\frac{\sigma_j}{\sqrt{2}}\;\;
    &\;\;\frac{\sigma_j}{\sqrt{2}}\;\;
    &\;\;\frac{\sigma_j}{\sqrt{2}}\;\;
    \\
    \begin{array}{c}
      \includegraphics[scale=0.7]{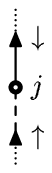} 
    \end{array}
    &\;\;\frac{-\Exp^{-\ImgI\gamma_j}}{\sqrt{2}}\;\;
    &\;\;\frac{\ImgI \sigma_j}{\sqrt{2}}\;\;
    &\;\;\frac{\Exp^{-\ImgI\beta_j}}{\sqrt{2}}\;\;
    &\;\;\frac{\Exp^{-\ImgI\beta_j}}{\sqrt{2}}\;\;
    &\;\;\frac{\Exp^{-\ImgI\beta_j}}{\sqrt{2}}\;\;\\
    \begin{array}{c}
      \includegraphics[scale=0.7]{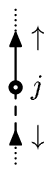} 
    \end{array}
    &\;\;\frac{\Exp^{\ImgI\gamma_j}}{\sqrt{2}}\;\;
    &\;\;\frac{\ImgI \sigma_j}{\sqrt{2}}\;\;
    &\;\;\frac{\Exp^{\ImgI\beta_j}}{\sqrt{2}}\;\;
    &\;\;\frac{\Exp^{\ImgI\beta_j}}{\sqrt{2}}\;\;
    &\;\;\frac{\Exp^{\ImgI\beta_j}}{\sqrt{2}}\;\;\\
    \begin{array}{c}
      \includegraphics[scale=0.7]{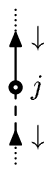} 
    \end{array}
    &\;\;\frac{\Exp^{-\ImgI\beta_j}}{\sqrt{2}}\;\;
    &\;\;\frac{\Exp^{-\ImgI\beta_j}}{\sqrt{2}}\;\;
    &\;\;\frac{-\sigma_j}{\sqrt{2}}\;\;
    &\;\;\frac{-\sigma_j}{\sqrt{2}}\;\;
    &\;\;\frac{-\sigma_j}{\sqrt{2}},\;\;\\
  \end{array}
  \label{eq:dvertices}
\end{equation}
where $0\le\gamma_j,\beta_j<2\pi$ are random phases
and $\sigma_j,\tau_j=\pm 1$ with equal probability.

\subsection{\label{sec:expansion} Diagrammatic expansion of the
ensemble averaged form factors}

For the Wigner-Dyson classes we are only interested in the
second-order form factor as the first-order form factor vanishes
exactly under the ensemble average. For the graphs
in the novel classes the ensemble average is non-trivial for
the first-order form factor. The second-order form
factor in the novel classes would contain additional
contributions proportional to $\langle s_n^2 \rangle$
and its complex conjugate which vanish in the Wigner-Dyson case.
We will not consider the second-order form factor for other
classes than the Wigner-Dyson here. 

Though the second-order form factor is a sum over pairs
of periodic orbits and the first-order form-factor
contains a single orbit the diagrammatic expansion is based
on similar observations. 
We will start with the Wigner-Dyson case.
Most pairs of periodic orbits
do not survive the ensemble average. A contribution can
only survive if all d-vertices have a partner d-vertex
such that the product of their factors does not depend
on the random phases or random signs. In the 
This condition can only be fulfilled if the two
orbits visit the same peripheral vertices with the
same multiplicities. The order may however be different and
one may introduce diagrams to denote the various appearing
permutations.  
Let us start with the
diagrams for class  $A$I and later
introduce the spin degrees of freedom.
An example of such a diagram for $n=6$ is given by
\begin{equation} 
  \begin{split}  
    \scriptstyle D=&
    \begin{array}{c}
      \includegraphics[scale=0.8]{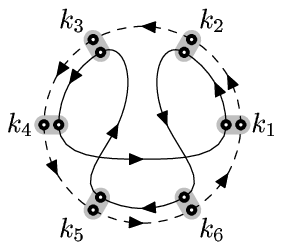}
    \end{array}\\ 
    \scriptstyle =& \scriptstyle 
    \xi \overset{B}{\underset{k_1,\dots,k_6=1}{\sum}} 
    \frac{\Exp^{\frac{2\pi\ImgI}{B}
        \overset{6}{\underset{m=1}{\sum}}
        (k_{\tilde{\pi}(m)}k_{\tilde{\pi}(m+1)}-k_mk_{m+1})}}{B^6}=
    \frac{\xi}{B^2}.
  \end{split} 
  \label{eq:ex_diagram}
\end{equation} 
In this diagram d-vertices have
been joined to pairs in a grey area to indicate that they carry 
the same
index $k_m$. 
We will call these grey areas \emph{scattering regions}
and, in the sequel, we will drop the indices
$k_m$ (as well as line indices for multi-component wave functions).
The \emph{multiplicity factor} $\xi$ will be explained later in this
section.
  
Any  diagram that contributes to $K_{2,n}$
has $2n$ d-vertices connected by $2n$ (directed) lines that define
two periodic orbits of length $n$. One of the orbits has
only full lines the other only dashed lines.
Each d-vertex and each line
contributes with the corresponding factor to the
diagram -- since by construction
the phase factors $\Exp^{\ImgI \beta_j}$ each have a partner 
$\Exp^{-\ImgI \beta_j}$ the phases all cancel.  
The number $w$ of scattering regions
may range in $w=n,n-1,\dots,1$. 
Each of the $w$ scattering regions
carries a single index $k_m$ ($m=1,2,\dots,w$)
for all d-vertices which it contains.
The number of d-vertices in a scattering region is always
even -- half of the d-vertices are part of each of
the two periodic orbits. 
If $w<n$ we will call the diagram a \emph{sub-diagram} -- in sub-diagrams
some g-vertices contain more than two d-vertices.

In the classes $A$ and $A$II
each line gets an additional index $\alpha_j$ 
($j=1,2,\dots,2n$) for the different (spin and iso-spin) components
of the wave function. 
The sum over $\alpha_j$ collapses
to a sum over allowed component configurations 
when under the averages over $\delta_j$ and $\gamma_j$.
An allowed component configuration is a set of line indices
$\alpha_j$ for which all phases  $\delta_j$ and $\gamma_j$
along the diagram cancel exactly. Then the product
of all phase factors is $\pm 1$.
The sum over the $w$ indices $k_m$ for
the different encounter regions and the sum over allowed
component configurations 
factorizes
such that the value of a diagram $D_{\nu}$ 
falls into three parts 
\begin{equation}
  D_{\nu}=\xi_{\nu} C_{\nu} P_{\nu}. 
\end{equation}
Here, $\xi_nu$ is the multiplicity factor that, $C_{\nu}$
is the \emph{quasi-spin} factor, and $P_\nu$
the \emph{principal part}.
The latter  contains only the line factors 
$\frac{1}{\sqrt{B}}\Exp^{\pm \ImgI\frac{2\pi}{B}k_{m}k_{m'}}$ and 
is  summed over the 
$w$ scattering region indices $k_m$. 
The quasi-spin factor $C_\nu$
contains all the d-vertex factors
which are summed over all allowed quasi-spin (component)
configurations -- in
class $A$I one has $C_\nu=1$ in the classes $A$ and $A$II
it is given by $\pm \frac{1}{2^n}$ 
for each allowed configuration. 
In class $A$ the sign is always positive, so $C_\nu$ is
$\frac{1}{2^n}$ times the number of allowed configurations.
In class $A$II quasi-spin configurations a negative sign
appears if an allowed
configuration contains an odd number of $d$-vertices where
the incoming spins and iso-spins are anti-parallel and both flip
(see \eqref{eq:vert_AII}).

Finally, the multiplicity factor $\xi_\nu=\frac{\tilde{\xi}}{n}$ 
is the number $\tilde{\xi}$
of times the diagram appears as a sub-sum in the original
form factor \eqref{eq:AIff}
before the average was performed divided by the length of
the orbit $n$. 
In general the sum over the indices $k_m$ is a sum over pairs of
points on two different periodic orbits. However, any
pair of two points along the same orbits will give exactly the same 
contribution. Thus, in general the sum over $k_m$ appears
$\tilde{\xi}=n^2$ times in the original one -- 
in the diagram \eqref{eq:ex_diagram}
one has indeed $\tilde{\xi}=n^2=36$ times. 
There are exceptions whenever one orbit is a repetition of a shorter
orbit or if the diagram is invariant with respect to some cyclic
permutation of the indices $k_m$. For example the diagrams
\begin{equation}
  D_{1}^{(0)}=
  \begin{array}{c}
    \includegraphics[scale=0.7]{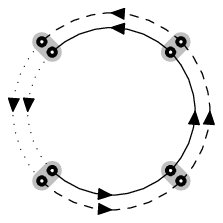}
  \end{array}
  \;\; \text{and}\;\;
  D_{2}^{(0)}=
  \begin{array}{c}
    \includegraphics[scale=0.7]{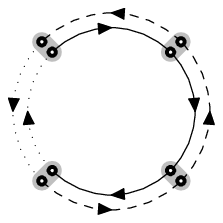}
  \end{array}
  \label{eq:AIdiag1}
\end{equation}
only appear $n$ times in the original sum such that $\tilde{\xi}=n$
and $\xi_1^{(0)}=\xi_2^{(0)}=1$. These are the two diagrams of
the diagonal approximation to be discussed in the next section. 

Obviously every pair of periodic orbits visiting the 
same scattering regions defines a diagram. 
However, if one sums over all the scattering
region indices $k_m$ without restriction
the same pair of periodic orbits may appear in different diagrams and
we have to face the problem of double (or multiple)
counting of periodic orbits. One way to get rid
of the double counting problem is to restrict the sum over
$k_m$ such that $k_i\neq k_j$ for $i\neq j$ and add 
all sub-diagrams
where the number $w$ of scattering regions is smaller than $n$
(with the same restriction in the sum over indices). 
The form factor
$K_{2,n}$ can then be written as a sum over all diagrams 
with $w=n,n-1,\dots,1$ scattering regions
and every pair of periodic orbits 
is counted exactly once.

It will be more convenient for us to keep
an unrestricted sum over the scattering regions 
and subtract the multiply counted counted orbits.
Any diagram $D$  
contains many sub-diagrams which can be obtained from $D$ by 
combining some scattering regions to
a single one.
This is equivalent
to restricting the sum in the diagram to $k_i=k_j$. Multiple counting
occurs when either two (or more) diagrams $D_1$ and $D_2$ have the same
sub-diagram or when one sub-diagram appears more than once
in the same diagram $D$. 
If the wave function has one component we just have to
subtract the overcounted sub-diagrams. In the presence
more than one
component  only the overcounted 
quasi-spin configurations 
have to subtracted. It may happen that a sub-diagram allows 
new quasi-spin configurations that have not been counted in the
original diagram -- then the corresponding configurations have to
be added. Luckily we will not encounter such difficulties
in the sequel.

Finally, we
may write the second-order form factor as a sum over diagrams
\begin{equation}
  K_{2,n}=\frac{1}{g\mu B}\langle |s_n|^2\rangle=
  \frac{n}{g \mu B}\left(\sum_{\nu} D_{\nu}- \sum_{\nu'} D^{\text{sub}}_{\nu'}
  \right).
  \label{eq:sumdiagrams}
\end{equation} 
The sum over diagrams $D_{\nu}$ only contains diagrams with $w=n$ 
scattering regions.
The sum over sub-diagrams $D^{\text{sub}}_{\nu'}$ accounts for 
the corrections due to multiple 
counting -- it contains diagrams with $w=n-1,n-2,\dots,1$ 
scattering regions and we include the number
of times it has been overcounted in the multiplicity factor. 
Note, the
factor $n$ outside the parentheses -- this factor appears due
to the definition of the multiplicity factor $\xi$. 

The diagrammatic representation of the first-order form factor
in the novel symmetry classes is analogous. The main difference is that
there is only a single periodic orbit connecting $n$ d-vertices any diagram
that contributes to $K_{1,n}$. Due to complete Andreev scattering at the
peripheral vertices.
As a consequence the d-vertices in the diagrams
always connect a full line with a dashed line.  
Diagrams can thus only
be drawn if the length $n$ of the orbit is even and
\begin{equation}
  s_n=0 \qquad \text{if $n$ is odd}.
\end{equation}
Similar as before for the second-order form factor most
contributions to the trace $s_n$ do not survive the average over the phases 
$\gamma_j,\beta_j$
(and signs $\sigma_j,\tau_j$).
The non-vanishing contributions can again be grouped
in diagrams where at most $\frac{n}{2}$ d-vertex indices remain
independent.
In a non-vanishing diagram each d-vertex has again a partner
such that the product of their d-vertex factors does not depend on
the random phase factors or signs. We again introduce
\emph{Scattering regions} that are
defined as in the diagrams to the second-order form factor.
Each scattering region 
has a single index which is the same for all d-vertices it contains.
A scattering region always contains an even number of d-vertices.

A general diagram will be drawn without indices and  
has a value $\hat{D}_\nu=\hat{\xi}_\nu \hat{C}_\nu \hat{P}_\nu$ where
the definitions of the principal part $\hat{P}_\nu$ and the quasi-spin factor
$\hat{C}_\nu$ are as before. The hat serves as a symbol to distinguish
between the contributions to the first-order
form factor (with hat) form those to the second-order form factor (no hat).
The principal part is independent of the 
symmetry group and given by the line contributions summed
over the indices of the scattering regions.

The quasi-spin factor $\hat{C}_\nu$
is the sum of the d-vertex factors over allowed spin and electron-hole
configurations. 
The quasi-spin factor may also vanish. If there is no spin it
will be $\pm 2$ where the factor $2$ is due to interchanging all electron and
hole lines.

The multiplicity factor $\hat{\xi}$ is here defined as the number
of times an equivalent sum appears in the original
sum $s_n$ -- note, that here we have not divided this number by $n$
as in the case of the diagrams for the second-order form factor. 
As an example the two diagrams
\begin{equation}
  \hat{D}_{1}^{(0)}=\begin{array}{c}
    \includegraphics[scale=0.7]{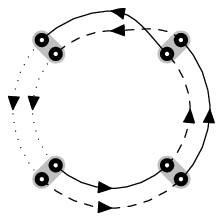}
  \end{array}
  \;\; \text{and}\;\;
  \hat{D}_{2}^{(0)}=
  \begin{array}{c}
    \includegraphics[scale=0.7]{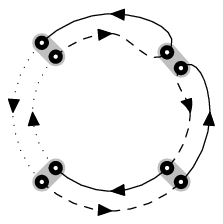}
  \end{array}
  \label{eq:selfdual}
\end{equation}
have multiplicity factors $\hat{\xi}_1^{(0)}=1$ and 
$\hat{\xi}_2^{(0)}=\frac{n}{2}$.
The difference is due to the different symmetry in the two diagrams. 
For the second
diagram the rotational symmetry is broken by the turning point of the
periodic orbit. These two diagrams correspond to the self-dual approximation
to be discussed in the next section.

Multiple countings have to be accounted for in the same manner as for the
second-order form factor which results in
\begin{equation}
  K_{1,n}=\frac{2}{g}\langle s_n \rangle
  =\frac{2}{g}\left(\sum_{\nu}D_\nu- \sum_{\nu'} D^\text{sub}_{\nu'}\right)
\end{equation}
by including the corresponding
sub-diagrams. 
 For finite $n$ the sum over diagrams is finite for both types
of form factors and the expansion converges
absolutely. As every periodic orbit (pair of period orbits) 
defines some diagram 
no contribution is neglected and the expansion is formally
exact. 

\section{\label{sec:short-time} The diagonal and self-dual approximations 
  and beyond}

While the calculation of the principal part, quasi-spin factor and
multiplicity factor for any given diagram is quite simple, the sum
over all diagrams defined in the previous
section is quite non-trivial. 
We will now give a systematic short-time expansion of this sum in the
ergodic limit $B\rightarrow \infty$.
In the leading order only two diagrams have to
be accounted for in each symmetry class.
More diagrams have to be taken into account for the next to leading
order where multiple-counting of periodic orbits lead to additional
complexity. 

\subsection{\label{sec:WD_short_time} 
  The diagrammatic short-time expansion of the form factors}

We will be interested in the short-time behavior of large
graphs such that we may assume $1\ll n \ll \frac{\mu B}{g}$.  
The first inequality $1\ll n$
assures that we are in the universal regime -- the ultra-short-time
behavior where $n=\mathcal{O}(1)$ is known to be dominated by the
system-dependent shortest orbits.  The second inequality can be
rewritten as $\tau\equiv \frac{g n}{\mu B}\ll 1$ which shows that we are
interested in times much shorter than Heisenberg time. 

We want to calculate $K_{1,n}$ and
$K_{2,n}$ in the limit $n,B\rightarrow \infty$
where $\tau=\frac{g n}{\mu B}\ll 1$ is constant. 
For that aim we will expand
the form factors in orders of $\tau$:
\begin{align} 
  K_1(\tau=\frac{gn}{\mu B})=& \overline{\frac{2}{g} \langle s_n \rangle}=&
  \frac{2}{g} \sum_{m=0}^\infty \mathcal{K}_{1}^{(m)}\\
  K_{2}(\tau=\frac{gn}{\mu B})=&\overline{\frac{1}{g \mu B}
    \langle |s_n|^2\rangle}=&
  \frac{\tau}{g^2} \sum_{m=0}^\infty \mathcal{K}_2^{(m)}
\end{align} 
where $\mathcal{K}_{1,2}^{(m)}$ are 
expansion coefficients proportional to $\tau^m$
that we have to calculate -- 
in this expansion we have anticipated that the leading
order will be a constant for $K_1(\tau)$ and will be linear in $K_2(\tau)$.
  
Each of these coefficients is a 
time averaged sum over some diagrams (and sub-diagrams).
The time average is performed over a short discrete time interval $[n,n+\Delta n]$
and has to be performed before we take the limit $n,B\rightarrow \infty$.
The interval has to chosen such that
$\Delta \tau =\frac{g \Delta n}{\mu B}$ vanishes in that limit.

Let us shortly summarize the procedure: \textit{i.} find all diagrams
and sub-diagrams that contribute to 
a given order $\mathcal{K}_{1,2}^{(m)}$ for
finite $B\gg n \gg 1 $, \textit{ii.} for every sub-diagram with $w$
vertices count how often it appears as a sub-diagram of other (sub)diagrams
in the same order $\mathcal{K}_{1,2}^{(m)}$ \emph{and} in all smaller orders
$\mathcal{K}_{1,2}^{(m-1)},\dots$, \textit{iii.}
calculate the values of the diagrams and add them and
subtract the overcounted sub-diagrams, \textit{iv.} 
take the time average over $n$, \textit{vi} finally take the limit
$n,B \rightarrow \infty$. Obviously the procedure is recursive 
and one has to start with $m=0$.

The rest of this section will be devoted to the description  of the diagrams
that contribute to a given order $\mathcal{K}_{1,2}^{(m)}$. 
Since $\tau=\frac{g n}{\mu B}$
this is equivalent to finding the diagrams with values of order $B^{-m}$.
A single diagram which is of order $B^{-m}$ in $B$ may have a very different
order in $n$ -- such that the limit $n,B \rightarrow \infty$ cannot
always be performed for a single diagram. Our procedure
will be self-consistent if this limit exists \emph{after}
we have summed over all (sub)diagrams.

As the multiplicity factor and the quasi-spin factor
do not depend on $B$ we have to look at the principal part which does
not depend on the symmetry class. The resulting expansion for the first-order
form factor can used for all graphs in the novel symmetry classes while
the diagrams in the expansion of the second-order form factor are
the same for all three Wigner-Dyson graphs. The difference between
the ensembles is mainly due to different quasi-spin factors. Note, that these
may vanish. 

The principal par of a (sub)diagram $D_{\nu}$ for $\mathcal{K}_{2}^{(m)}$  
($\hat{D}_{\nu}$ for $\mathcal{K}_{1}^{(m)}$)
with $2n$ ($n$) d-vertices and lines and
$w\le n$ ($w\le \frac{n}{2}$) different scattering
regions is bounded from above
\begin{equation}
  P_{\nu}\le \frac{1}{B^{n-w}}\qquad \hat{P}_{\nu}\le \frac{1}{B^{\frac{n}{2}-w}}\ .
  \label{eq:principle1}
\end{equation}  
Indeed, the absolute value of the summand
in the principal part is $\frac{1}{B^n}$ ($\frac{1}{B^\frac{n}{2}}$) 
stemming from the $2n$ ($n$)
amplitudes of the line contributions when one sums over
$w$ indices $k_m=1,2,\dots,B$. In \eqref{eq:principle1} 
equality holds if all the phases 
acquired along the lines cancel exactly. This is the case
in \emph{complete} (sub)diagrams for which
every full line that connects two scattering
regions is accompanied by
a dashed line connecting the same two scattering regions.
We will call such a pair of lines a \emph{complete diagonal (anti-diagonal)} 
pair of lines if they start and end at the same (opposite) scattering region.

Every complete (sub)diagram with $w=n-m$ ($w=\frac{n}{2}-m$)
scattering regions contributes as a (sub)diagram to $\mathcal{K}_{2}^{(m)}$
($\mathcal{K}_{1}^{(m)}$). If $w<n-m$ ($w<\frac{n}{2}-m$) no sub-diagrams
exist that contribute. 

It remains to find the
non-complete (sub)diagrams. For non-complete diagrams the sum is oscillating
due to the appearing phase factors. 
An oscillating sum over one index is of the form
\begin{equation}
  \sum_{k=1}^B \Exp^{\frac{2\pi\ImgI}{B}\, k(k'-k'')}= B \delta_{k'k''}
\end{equation}
where $\delta_{k'k''}$ is the Kronecker symbol. The subsequent sum over $k'$
does not give an additional factor $B$.
Thus non-complete (sub)diagrams can only contribute to
$\mathcal{K}_{2}^{(m)}$ ($\mathcal{K}_{1}^{(m)}$) if they have a 
complete sub-diagram with $w=n-m$
scattering regions.

\emph{All} (sub)diagrams for $\mathcal{K}_{2}^{(m)}$ 
($\mathcal{K}_{1}^{(m)}$) are thus found by
first finding all complete (sub)diagrams with $n-m$ ($\frac{n}{2}-m$)
scattering regions
and then finding all (non-complete) diagrams with $n-m+1,\dots,n$ 
($\frac{n}{2}-m+1,\dots,\frac{n}{2}$)
scattering regions which contain one of the complete sub-diagrams.

\subsection{\label{sec:diagonal_selfdual} 
  The diagonal approximation and beyond for star graphs 
  in the Wigner-Dyson classes}

We will now find all diagrams that contribute to the
diagonal approximation ($m=0$) and one order beyond in the
second-order form factor of star graphs in the Wigner-Dyson classes.
The coefficients $\mathcal{K }_2^{(0)}$ and $\mathcal{K}_2^{(1)}$
will be calculated. The resulting form factors will be in accordance
with the random-matrix predictions \cite{usI} (for $\tau\ll 1$) 
\begin{equation}
  K_2(\tau)=
  \begin{cases}
    \tau & \text{GUE ($A$-GE)}\\
    2\tau-2\tau^2+\mathcal{O}(\tau^3) & \text{GOE ($A$I-GE)}\\
    \frac{\tau}{2}+\frac{\tau^2}{4}+\mathcal{O}(\tau^3) & \text{GSE ($A$Ii-GE)}.
  \end{cases}
  \label{eq:WD_RMT}
\end{equation}

\subsubsection{\label{sec:WD_diagrams} The diagrams}

For the diagonal approximation ( $m=0$) -- we just have to find all 
complete diagrams with two periodic orbits of length $n$ and $w=n$ 
scattering regions.
There are two such diagrams: $D^{(0)}_1$ and $D^{(0)}_2$ given
by \eqref{eq:AIdiag1}. In $D^{(0)}_1$ the two periodic orbits are the same
and in  $D^{(0)}_2$ one orbit is the time-reversed orbit of the other.
We have
\begin{equation}
  \mathcal{K}_{2}^{(0)}=\overline{D^{(0)}_1+D^{(0)}_2}
\end{equation}
which gives the diagonal 
approximation. The multiplicities of the two diagrams are   
$\xi^{(0)}_1=\xi^{(0)}_1=1$ 
and their principal parts $P^{(0)}_1=P^{(0)}_2=1$.
Only the quasi-spin contributions depend
on the symmetry class such that
\begin{equation}
  D^{(0)}_1=C^{(0)}_1 \qquad \text{and} 
  \qquad D^{(0)}_2=C^{(0)}_2
\end{equation}
We will see in section \ref{sec:WignerDysonDiagonal}
that for broken time-reversal the quasi-spin contribution of the
diagram $D^{(0)}_2$
vanishes in the limit $n\rightarrow\infty$.

For the first order beyond the diagonal approximation 
($m=1$) we have to find all 
complete sub-diagrams with $n-1$ scattering regions
and then all new non-complete diagrams with $n$ scattering regions that 
contain one of the complete sub-diagrams. 
The diagrams can be grouped into three families. Each of the families contains
complete and non-complete diagrams.  
The complete diagrams of the first two families 
appear trivially as a sub-diagram
of the diagonal diagrams $D^{(0)}_1$ and $D^{(0)}_2$ 
by joining two of their scattering regions
\begin{subequations}
  \begin{align}
    D^{\text{sub},(1)}_{1,l}=&
    \begin{array}{c}
      \includegraphics[scale=0.7]{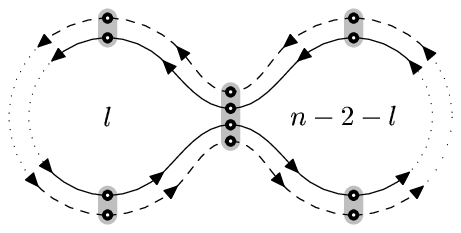}
    \end{array}&&=
    C^{\text{sub},(1)}_{1,l} \frac{n}{B}
    \label{eq:WDbeyond1}
    \\
    D^{\text{sub},(1)}_{2,l}=&
    \begin{array}{c}
      \includegraphics[scale=0.7]{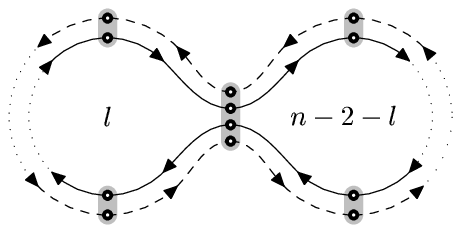}
    \end{array}&&= 
    C^{\text{sub},(1)}_{2,l} \frac{n}{B}
    \label{eq:WDbeyond2}
  \end{align}
\end{subequations}
where we have already included the principal part $P=\frac{1}{B}$
and the multiplicity $\xi=n$.
The integer $l=0,1,\dots,n-2$ gives the
number of scattering regions
in the left loop (the central scattering region is not counted).  
The diagram with $l$ vertices in
the left loop is equivalent to the one with $l$ vertices in the right
loop. 
The non-complete diagrams of the first two families
each contain one of corresponding complete sub-diagrams
\begin{subequations}
  \begin{align}
    D^{(1)}_{1,l}=&&
    \begin{array}{c}
      \includegraphics[scale=0.7]{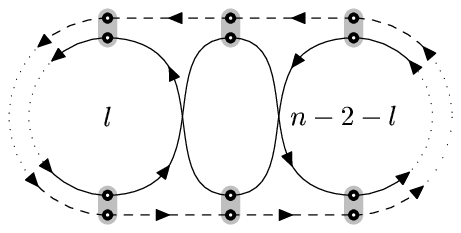} 
    \end{array}=&&
    C^{(1)}_{1,l}\frac{n}{B}\label{eq:WDbeyond4}\\
    D^{(1)}_{2,l}=&&
    \begin{array}{c} 
      \includegraphics[scale=0.7]{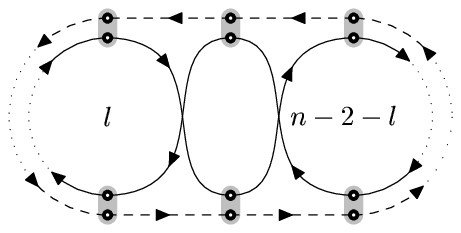}
    \end{array}=&&
    C^{(1)}_{2,l} \frac{n}{B}.\label{eq:WDbeyond5}
  \end{align}
\end{subequations} 
Again, $l=0,1,\dots$ is the number of scattering regions in the left loop
(the two central scattering regions are not counted).
Let us now assume that the quasi-spin factors within one family of
diagrams is the same $C_{1,l}^{(1)}=C_{1,l}^{\text{sub},(1)}\equiv C_1^{(1)}$
and $C_{1,l}^{(1)}=C_{2,l}^{\text{sub},(1)}\equiv C_2^{(1)}$. 
In the limit $n,B \rightarrow \infty$ this can indeed be shown for each of
the symmetry classes.
The contribution
of these two families then vanishes due to overcounting. Each of the sub-diagrams
has been counted once in the diagonal approximation and is also a sub-diagram
of the new non-complete diagrams. 

The third family contains all non-trivial diagrams that correspond
to the Sieber-Richer pairs.
They consist of
two loops as well, however in one loop the two
orbits are parallel while in the other they are time-reversed
\begin{equation}
  D^{\text{sub},(1)}_{3,l}=
  \begin{array}{c}
    \includegraphics[scale=0.7]{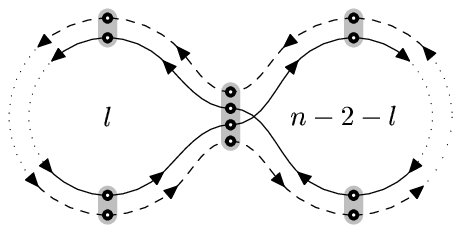}
  \end{array}
  =C^{\text{sub},(1)}_{3,l}
  \frac{n}{B}.
  \label{eq:WDbeyond3}
\end{equation}
In these diagram it is irrelevant if we draw the crossing on the left
or on the right of the central scattering region. 
If the left or the right loop has either zero or one vertex it is
indistinguishable from one of the corresponding
previous diagrams
\eqref{eq:WDbeyond1} or \eqref{eq:WDbeyond2}. 
Thus $l=2,3,\dots,n-4$
and there are $n-5$ new diagrams of this form.
Note, that these diagrams are \emph{not} sub-diagrams of the diagonal diagrams
$D^{(0)}_1$ and $D^{(0)}_2$.

The corresponding non-complete diagrams diagrams are
\begin{equation} 
  D^{(1)}_{3,l}=\begin{array}{c} 
    \includegraphics[scale=0.7]{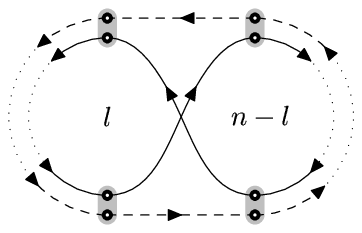}
  \end{array}=
  C^{(1)}_{3,l}\frac{n}{B}.
  \label{eq:WDbeyond6}
\end{equation}
The number of vertices in the left loop (left of the crossing)
runs from $l=4$ to $l=n-4$ --
all other diagrams are indistinguishable from 
corresponding diagrams in the families
\eqref{eq:WDbeyond4} or \eqref{eq:WDbeyond5}. 
Thus there are $n-7$ new diagrams in this family.

Each of the sub-diagrams $D^{\text{sub},(1)}_{3,l}$ is contained
once
in the diagrams $D^{(1)}_{3,l}$ and $D^{(1)}_{3,l+2}$ if $4\le l\le n-6$
while for $l=2,3$ ($l=n-4,n-5$) they are contained once
in one $D^{(1)}_{2,l}$ and $D^{(1)}_{3,l+2}$
($D^{(1)}_{1,l+2}$ and $D^{(1)}_{3,l}$). Thus each 
sub-diagrams $D^{\text{sub},(1)}_{3,l}$ has been overcounted once.
Again one can show that $C_{3,l}^{(1)}=C_{3,l}^{\text{sub},(1)}\equiv C_3^{(1)}$
in the limit $n,B \rightarrow \infty$.
The contributions does then not vanish if $C_3^{(1)}\neq 0$ and one
has
\begin{equation}
  K_2^{(1)}=-2 \overline{C^{(1)}_3}.
\end{equation} 

\subsubsection{\label{sec:WignerDysonDiagonal}
  Wigner-Dyson class $A$I (GOE)}

Since the graphs in symmetry class $A$I have a one-component wave 
function ($\mu=g=1$) 
the quasi-spin contribution to any (sub)diagram $D_{\nu}$ is 
$C_{\nu}=1$. In the diagonal approximation each the two diagrams has the value
$D^{(0)}_{1}=D^{(0)}_{2}=1$ and we have 
\begin{equation}
  \mathcal{K}_2^{(0)}=2.
\end{equation}
For the next order we have $D^{(1)}_{3,l}=D^{\text{sub},(1)}=\tau$
which gives
\begin{equation}
  \mathcal{K}^{(1)}_2=-2\tau.
\end{equation} 
Altogether the leading terms of the second-order form factor give
$K_2(\tau)=2\tau-2\tau^2 
  +\mathcal{O}(\tau^3)$ in accordance with the GOE prediction 
\eqref{eq:WD_RMT}.

\subsubsection{\label{sec:WignerDysonDiagonalA}
  Wigner-Dyson class $A$ (GUE)}

In class $A$ we have defined star graphs with a two-component wave
function ($\mu=2$, $g=1$) and 
we have to calculate the quasi-spin contributions for all
diagrams. 

Let us start with diagram $D^{(0)}_1$ where
both periodic orbits are the same (parallel). Only spin configurations
survive the average over the phases $\gamma_k$ and $\delta_k$
for which the spins on two parallel lines is the same. The spins
on lines that connect different scattering regions are independent. Thus
there are $2^n$ allowed spin configurations and the quasi-spin contribution to
the diagram is  $C^{(0)}_1=1$ which results in 
\begin{equation}
  D^{(0)}_1=1.
\end{equation} 

In the diagram $D^{(0)}_2$ the two orbits are anti-parallel. 
Here, the spins on different lines are not independent
for configurations that survive the phase average. 
If at any pair of anti-parallel lines
the two spins are $\sigma_1$ for the full line
and $\sigma_2$ for the dashed line
the spins on the neighboring lines
are $\sigma_1$ for the dashed line and $\sigma_2$ for the full line.
If $n$ is even $\sigma_1$ and $sigma_2$ are independent
while for odd $n$ they are equal.
So there are only four (two) allowed spin configurations 
for even (odd) $n$ which gives
\begin{equation}
  D^{(0)}_2=\frac{3+(-1)^{n}}{2^{n}}.
\end{equation}
This value is exponentially suppressed in the limit $n\rightarrow \infty$ --
a consequence of
breaking the time-reversal invariance in this symmetry class.  

In the first order beyond the diagonal approximation the contribution of
the first family vanishes because the quasi-spin factor is the same throughout
the family (see section \ref{sec:WD_diagrams}). 
The contribution of the second family vanishes because their quasi-spin factor
all vanish in the limit $n,B \rightarrow \infty$.
For the third family the quasi-spin factors
are not equal for each diagram in the group. Its value is small
if it contains a big anti-parallel loop. There are however diagrams
with a short anti-parallel loop. It is however not difficult to
calculate the quasi-spin factor  of each diagram which gives
\begin{equation}
  \begin{array}{ll}
    D^{(1)}_{3,l}&=\frac{3-(-1)^{n-l}}{2^{n-l-1}}\frac{n}{B}\\
    D^{\text{sub},(1)}_{3,l}&=\frac{3-(-1)^{n-l}}{2^{n-l-1}}\frac{n}{B}
  \end{array}
\end{equation}
The sum over all these contributions (including the correct accounting for
multiple counting) is
\begin{equation}
  \scriptstyle
  \overset{n-4}{\underset{l=4}{\sum D^{(1)}_{3,l}}} -
  \overset{n-4}{\underset{l=2}{\sum D^{\text{sub},(1)}_{3,l}}}=
  -D^{\text{sub},(1)}_{3,2}-D^{\text{sub},(1)}_{3,3}=
  -\frac{9+(-1)^n}{2^{n-3}}\frac{n}{B}
\end{equation}
which also vanishes in the limit $n\rightarrow \infty$.
Thus Sieber-Richter pairs do not contribute for broken time-reversal
symmetry.
 
Only the diagrams of the diagonal approximation contribute
to the form factor which has
the value $K_2(\tau)=\tau$ as predicted by the GUE \eqref{eq:WD_RMT}
for $\tau<1$.

\subsubsection{\label{sec:WignerDysonDiagonalAII}
  Wigner-Dyson class $A$II (GSE)}

In class $A$II we have a four-component wave function on the
star graphs ($\mu=4$, $g=2$). 
At each d-vertex iso-spin flips while spin may either flip or not.
As an immediate consequence the length of every periodic orbit is even
and $s_{n}=0$ if $n$ is odd.

For the diagonal approximation we have to
recalculate the quasi-spin factors 
of the two diagrams $D^{(0)}_1$ and $D^{(0)}_2$ with the additional
spin and iso-spin freedoms. 
Let us start with the first diagram where
both orbits traverse the scattering regions
in the same order. The iso-spins
on parallel lines are either always parallel or always anti-parallel.
If iso-spins on parallel lines are parallel only the spin configurations 
which are everywhere parallel as well survive the average.
As the spins on lines connecting different scattering regions are independent
there are $2^{n+1}$ such configurations 
with parallel iso-spins ($n$ factors $2$ from the spins and
one factor from the iso-spin).  
 
If the iso-spins are all anti-parallel there are allowed configurations 
with either all spins parallel or all anti-parallel between two 
scattering regions.
If the spins are all anti-parallel they never flip and if they
are parallel they both flip at every scattering region.  
Altogether there are only $8$ configurations
with anti-parallel iso-spins which implies
that these contributions are negligible in the limit $n,B\rightarrow \infty$. 
Note, that anti-parallel iso-spins implies
that the two orbits are different and not related by time-reversal.
At each d-vertex with an incoming spin that is antiparallel
to the incoming iso-spin a factor $-1$ is gathered if the spin
flips. All factors $-1$ within one scattering region  cancel because the two
d-vertices have the same configuration.

Due to time-reversal 
invariance the two diagrams in the diagonal approximation have the same
value
\begin{equation}
  D^{(0)}_1=
  D^{(0)}_2=
  2+2^{-n+3}.
  \label{eq:AdiagII1}
\end{equation}
Note, that time-reversal implies changing the directions of arrows
flipping the spins while iso-spins do not flip. In the
time-reversed diagram $D^{(0)}_2$ the factors $-1$ cancel in a slightly
different way. As on neighboring lines spins are always anti-parallel
and iso-spins always parallel one gets a factor $-1$ at every 
scattering region where both spins flip. Since $n$ is even so is  
the number of spin flips along each orbit.

For the calculations of the diagrams that contribute beyond 
the diagonal approximation we will neglect all contributions that
are exponentially suppressed. The first two families have
a vanishing contribution as their quasi-spin factors are the same within each family 
in the limit $n,B \rightarrow \infty$. Again, we only have to consider
the Sieber-Richter family with
the diagrams $D^{(1)}_{3,l}$ and $D^{\text{sub},(1)}_{3,l}$. In both families
only even $l$ can give contributions that are not exponentially 
suppressed as iso spins have to remain parallel in the left and right loops.
There are $\frac{n}{2}-3$ 
contributing diagrams in the family  $D^{(1)}_{3,l}$ and
$\frac{n}{2}-2$ in the family $D^{\text{sub},(1)}_{3,l}$ (each of
them has been overcounted once). They all have the same value
\begin{equation}
  D^{(1)}_{3,l}=D^{\text{sub},(1)}_{3,l}=-\frac{n}{B}.
\end{equation}
For any quasi-spin configuration an odd number of factors $-1$
is gathered along the orbits -- also the spins on the
lines that connect the left and right loops are not independent.

Altogether we get
\begin{equation}
  K_{2,n}=\frac{n}{8 B}
  \begin{cases}
    0 & \text{if $n$ odd,}\\
    4 + \frac{n}{B} +\mathcal{O}(\frac{n^2}{B^2})
    &\text{if $n$ is even.}
  \end{cases}
\end{equation}
As $\tau=\frac{n}{2B}$ the time average yields
$K_2(\tau)=\frac{\tau}{2}+\frac{\tau^2}{4}
  +\mathcal{O}(\tau^2)$
in accordance with the short-time expansion of the universal result \eqref{eq:WD_RMT}
from the GSE . 

\subsection{The self-dual approximation and beyond for chiral and Andreev
  graphs}

Now we will consider the self-dual approximation ($m=0$) and
one order beyond ($m=1$) for the first-order form factors in the
novel symmetry classes. We will give all diagrams and show that they
add up to the corresponding random-matrix predictions \cite{usI}
for $\tau\ll 1$
\begin{equation}
  K_1(\tau)=
  \begin{cases}
    -1 & \text{$C$-GE}\\
    -1 +\frac{\tau}{2} +\mathcal{O}(\tau^2)& \text{$C$I-GE}\\
    1 & \text{$D$-GE}\\
    \frac{1}{2}-\frac{\tau}{8}+\mathcal{O}(\tau^2)& \text{$D$III-GE}\\
    -\frac{\tau}{2}+\mathcal{O}(\tau^2)& \text{chGUE($A$III-GE)}\\
    1-\frac{3 \tau}{2}+\mathcal{O}(\tau^2)& \text{chGOE($BD$I-GE)}\\
    -\frac{1}{2}-\frac{3 \tau}{8}+\mathcal{O}(\tau^2)& \text{chGSE($C$II-GE)}.
  \end{cases}
  \label{eq:novel_RMT}
\end{equation}

\subsubsection{The diagrams}

The self-dual approximation takes into account all complete
diagrams for $K_{1,n}$ (where $n$ is even) with $n/2$
scattering regions. 
The approximation has been called self-dual
because the diagrams contain those orbits which are invariant under
either a chiral symmetry or charge conjugation (in combination
with time-reversal).

There are two self-dual diagrams which
have been given in \eqref{eq:selfdual} where 
their multiplicity factors have been given. Their principal part 
is $\hat{P}^{(0)}_{1,2}=1$.
In the first diagram the same scattering regions
are visited twice in the same order
but with electrons replaced by holes in the second traversal. It
vanishes exactly if $\frac{n}{2}$ is even (no complete diagram
can then be drawn) while
\begin{equation}
  \hat{D}^{(0)}_1
  =\hat{C}^{(0)}_1,
  \label{eq:selfdual1}
\end{equation}
for odd $\frac{n}{2}$.
In the second diagram the same orbits are traversed in opposite (time-reversed)
direction. It contains one scattering region where the direction is changed. 
Its value is
\begin{equation} 
  \hat{D}^{(0)}_2
  =\frac{n}{2} \hat{C}^{(0)}_2.
  \label{eq:selfdual2}
\end{equation}
If $n/2$ is odd the turning point region has one incoming electron
and one incoming hole as drawn in \eqref{eq:selfdual}. The diagram has to 
be changed slightly for even $n/2$ -- then the turning point region has 
either two incoming electrons or two incoming holes. 

To calculate the linear order of the form factor we need
all families of
diagrams that contribute to $\mathcal{K}_{1}^{(1)}$.
Most of these diagrams will have a multiplicity factor $\hat{\xi}=\frac{n}{2}$
because there is no symmetry. The exceptions have $\hat{\xi}=\frac{n}{4}$
due to some two-fold symmetry. When we explicitly give the value of
a family of diagrams they always refer to the generic case where
no two-fold symmetry is present. The cumbersome accounting for all 
these cases will only be done for those families of diagrams that
do not vanish for different reasons.
 
The diagrams  have either $n/2$ scattering regions, are not complete
or they have $n/2-1$ scattering regions, are complete. 
In all cases  the principal value is $\hat{P}^{(1)}_\nu=\frac{1}{B}$. 
One may group the diagrams into seven different families.
 
The first two families have complete diagrams that are
sub-diagrams of the self-dual approximation.
Joining two scattering regions in the two diagrams \eqref{eq:selfdual}
gives
\begin{subequations}
  \begin{align}
    \hat{D}^{\text{sub},(1)}_{1,l}=&
    \begin{array}{c}
      \includegraphics[scale=0.7]{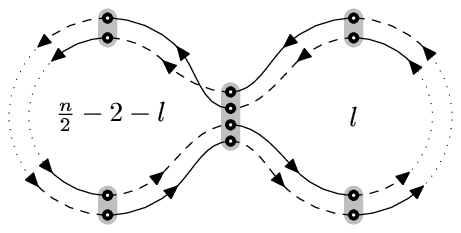} 
    \end{array}=\!\!&
    \hat{C}^{\text{sub},(1)}_{1,l}\frac{n}{2B}\label{eq:beyond_sd1}\\
    \hat{D}^{\text{sub},(1)}_{2,l,k}=&
    \begin{array}{c} 
      \includegraphics[scale=0.7]{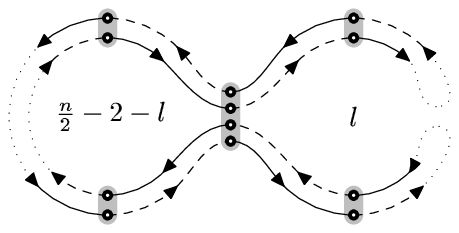}
    \end{array}=\!\!&
    \hat{C}^{\text{sub},(1)}_{2,l,k} \frac{n}{2B}.\label{eq:WDbeyond5}
  \end{align}
\end{subequations}
where $l$ is the number of scattering regions in the right loop.
The first type of diagrams only exists if
$\frac{n}{2}$ is odd.
For the second type of diagrams $k$ is an index for 
the different positions of the turning point.

There are two types of non-complete diagrams in each of the two families.
For the first family they are given by 
\begin{equation}
  \begin{array}{c}
    \hat{D}^{(1)}_{1a,l}=
    \begin{array}{c}
      \includegraphics[scale=0.7]{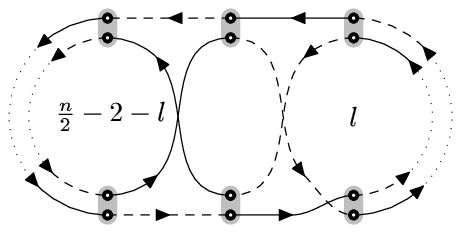}
    \end{array}
    =
    \frac{n}{2B} \hat{C}^{(1)}_{1a,l} 
    \\
    \hat{D}^{(1)}_{1b,l}=
    \begin{array}{c}
      \includegraphics[scale=0.7]{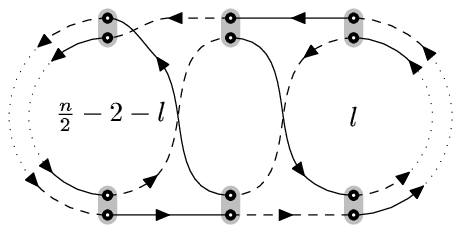}
    \end{array}
    =\frac{n}{2B} \hat{C}^{(1)}_{1b,l}.
  \end{array}
  \label{eq:beyond_sd1a}
\end{equation}
The difference between the two diagrams is that two central
scattering regions have two incoming lines of the same type for the first diagram
and two different incoming lines in the second. 
Both diagrams only exist 
for odd $\frac{n}{2}$. 

For the second family one has  
\begin{equation}
  \begin{array}{c} 
    \hat{D}^{(1)}_{2a,l,k}=
    \begin{array}{c}
      \includegraphics[scale=0.7]{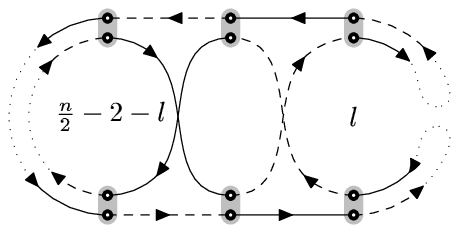} 
    \end{array}
    =
    \frac{n}{2B} \hat{C}^{(1)}_{2a,l,k} 
    \\
    \hat{D}^{(1)}_{2b,l,k}=
    \begin{array}{c}
      \includegraphics[scale=0.7]{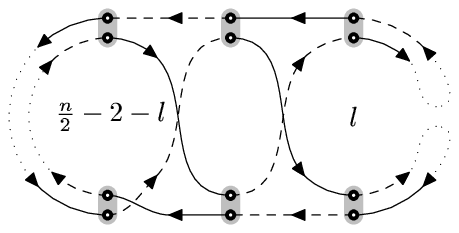} 
    \end{array}
    =\frac{n}{2B} \hat{C}^{(1)}_{2b,l,k}
  \end{array}
  \label{eq:beyond_sd2a}
\end{equation}
which differ in the direction of the lower line. Both
diagrams have been drawn for
an even number $\frac{n}{2}-l-2$ of scattering regions
in the left loop. Then the two central
scattering regions have one incoming dashed and one incoming full
line each. For an odd number
of scattering regions
in the left loop the central scattering regions have
two incoming lines of the same type. 

Each of the sub-diagrams
$D^{\text{sub},(1)}_{1,l}$ or $D^{\text{sub},(1)}_{2,l,k}$
has been counted three times. Once ion the self-dual approximation and
twice in the diagrams $D^{(1)}_{1a/1b,l,k}$ and $D^{(1)}_{2a/2b,l,k}$.
If the quasi-spin factors do not differ in the limit $n,B \rightarrow \infty$
the diagrams cancel due to overcounting. This is indeed the case for all seven
ensembles of star graphs in the novel symmetry classes.

The third family contains diagrams
with two parallel complete loops. The complete diagrams in this family 
\begin{equation}
  \hat{D}^{\text{sub},(1)}_{3,l}=
  \begin{array}{c}
    \includegraphics[scale=0.7]{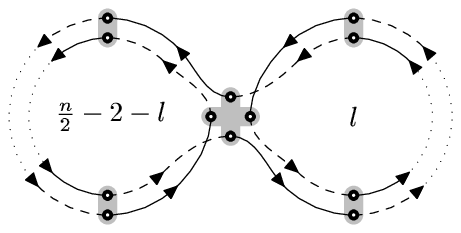}
  \end{array}.
  =
  \frac{n}{2B} \hat{C}^{\text{sub},(1)}_{3,l}
  \label{eq:beyond_sd3}
\end{equation}
cannot be obtained  as sub-diagrams of the self-dual diagrams.
In the second traversal of each loop the roles of electrons and holes
are interchanged. 
A complete diagram can only be achieved
if both $l=0,2,\dots$ and $\frac{n}{2}-l-2=0,2,\dots$ are even. 
Thus $\frac{n}{2}$
is even. The non-complete diagrams in this family are given by
\begin{equation}
  \begin{array}{rcl} 
    \hat{D}^{(1)}_{3a,l}=&
    \begin{array}{c}
      \includegraphics[scale=0.7]{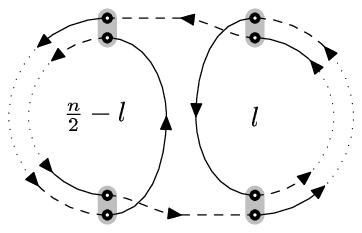} 
    \end{array}&
    =
    \frac{n}{2B} \hat{C}^{(1)}_{3a,l} 
    \\
    \hat{D}^{(1)}_{3b,l}=&
    \begin{array}{c}
      \includegraphics[scale=0.7]{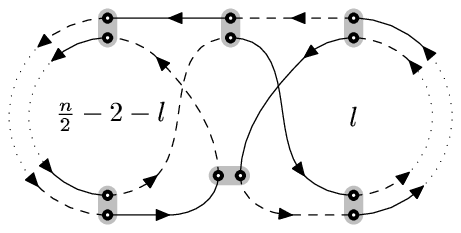} 
    \end{array}
    &
    =\frac{n}{2B} \hat{C}^{(1)}_{3b,l}.
  \end{array}
  \label{eq:beyond_sd3a}
\end{equation}
In both cases $\frac{n}{2}$ is even while and $l$ is odd in the first type
and even in the second type.
If all quasi-spin factors within this family are equal
the contribution of this family vanishes due to overcounting.
Indeed for fixed even $l$ the complete sub-diagrams
$D^{\text{sub},(1)}_{3,l}$ has been counted three times.
It appears once in $D^{(1)}_{3b,l}$
and twice in the diagram $D^{(1)}_{3a,l+1}$ 
because there are two ways of joining diagonally opposite scattering regions.
As the values of all diagrams are equal and
the complete  sub-diagram   has been overcounted twice for each fixed even $l$
the contributions cancel.
 
By introducing turning points in both loops or just in the right loop
of the third family one arrives at the fourth and fifth family
of diagrams.
The complete diagrams in the fourth family are given by
\begin{equation}
  \hat{D}^{\text{sub},(1)}_{4,l,k_l,k_r}=
  \begin{array}{c}
    \includegraphics[scale=0.7]{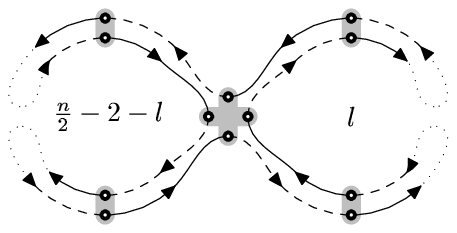}
  \end{array}
  =
  \frac{n}{2B} \hat{C}^{\text{sub},(1)}_{4,l,k_l,k_r}
  \label{eq:beyond_sd4}
\end{equation}
and the corresponding non-complete diagrams are
\begin{equation}
  \begin{array}{rcl} 
    \hat{D}^{(1)}_{4a,l,k_l,k_r}=&
    \begin{array}{c}
      \includegraphics[scale=0.7]{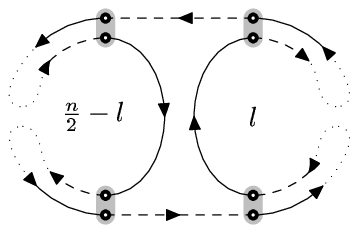} 
    \end{array}&
    =
    \frac{n}{2B} \hat{C}^{(1)}_{4a,l,k_l,k_r} 
    \\
    \hat{D}^{(1)}_{4b,l,k_l,k_r}=&
    \begin{array}{c}
      \includegraphics[scale=0.7]{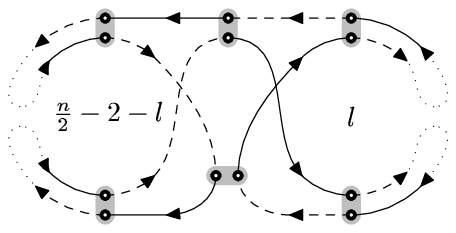} 
    \end{array}
    &
    =\frac{n}{2B} \hat{C}^{(1)}_{4b,l,k_l,k_r}.
  \end{array}
  \label{eq:beyond_sd4a}
\end{equation}
Here we need two indices $k_{l,r}$ to account for the positions of the two
turning points which makes counting quite cumbersome.
Luckily for fixed values of $k_{l,r}$ and $l$ one can show that
the contribution of this family vanishes
due to multiple counting if all quasi-spin factors are the same. The argument
is analogous to the third family.

The same argument also cancels
the contribution of the fifth family 
with the complete sub-diagrams 
\begin{equation}
  \hat{D}^{\text{sub},(1)}_{5,l,k}=
  \begin{array}{c}
    \includegraphics[scale=0.7]{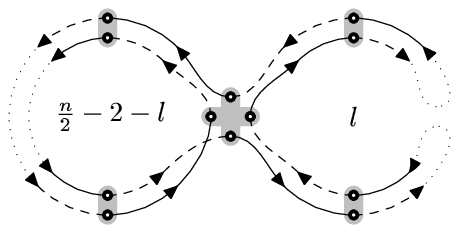}
  \end{array}
  =
  \frac{n}{2B} \hat{C}^{\text{sub},(1)}_{5,l,k}.
  \label{eq:beyond_sd5}
\end{equation}
In this diagram $\frac{n}{2}-l-2$ must be odd and the diagrams with
$1<l<\frac{n}{2}-3$ to give a new diagram. The corresponding
non-complete diagrams are
\begin{equation}
  \begin{array}{rcl} 
    \hat{D}^{(1)}_{5a,l,k}=&
    \begin{array}{c}
      \includegraphics[scale=0.7]{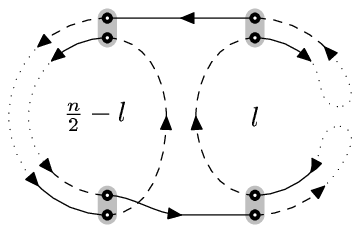} 
    \end{array}&
    =
    \frac{n}{2B} \hat{C}^{(1)}_{5a,l} 
    \\
    \hat{D}^{(1)}_{5b,l,k}=&
    \begin{array}{c}
      \includegraphics[scale=0.7]{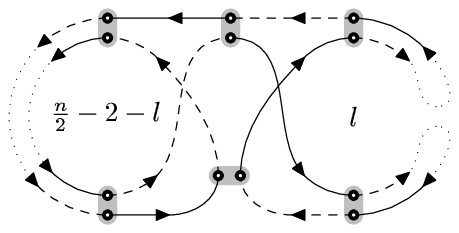} 
    \end{array}
    &
    =\frac{n}{2B} \hat{C}^{(1)}_{5b,l,k}.
  \end{array}
  \label{eq:beyond_sd5a}
\end{equation}

All non-trivial contributions to the first-order form factor 
come from the two remaining families of diagrams. 
They contain no turning point but do contain loops of
antiparallel lines. The complete diagrams of the sixth family
\begin{equation}
  \hat{D}^{\text{sub},(1)}_{6,l}=
  \begin{array}{c}
    \includegraphics[scale=0.7]{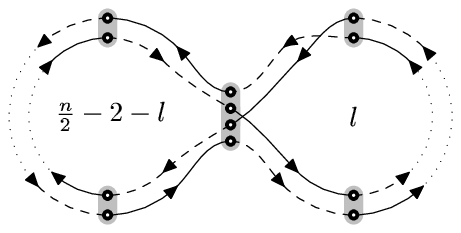}
  \end{array}
  =
  \frac{n}{2B} \hat{C}^{\text{sub},(1)}_{6,l}
  \label{eq:beyond_sd6}
\end{equation}
contain one parallel and one anti-parallel loop
while the complete diagrams
\begin{equation}
  \hat{D}^{\text{sub},(1)}_{7,l}=
  \begin{array}{c}
    \includegraphics[scale=0.7]{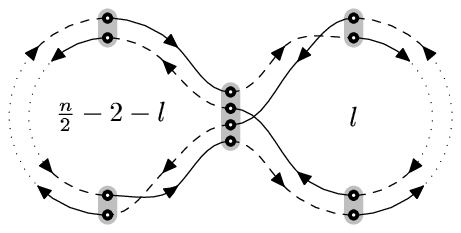}
  \end{array}
  =
  \frac{n}{2B} \hat{C}^{\text{sub},(1)}_{7,l}
  \label{eq:beyond_sd7}
\end{equation}
of the seventh family contain two anti-parallel
loops.
In both  types of complete diagrams $l$ is odd. In the sixth family
$\frac{n}{2}-2-l$ is even while
it is
odd in the seventh family. Thus the sixth family only exists for
odd $\frac{n}{2}$ and the seventh only for even $\frac{n}{2}$.
If $l=1$ the diagrams 
$D^{\text{sub},(1)}_{6,1}$ and $D^{\text{sub},(1)}_{7,1}$
are the same as $D^{\text{sub},(1)}_{2,1,k}$
(with $k$ such that the turning point is on the scattering
region  in the right loop).
For $l=\frac{n}{2}-2$ the diagram $D^{\text{sub},(1)}_{6,\frac{n}{2}-2}$
is the same as $D^{\text{sub},(1)}_{1,\frac{n}{2}-2}$. Thus for the sixth
family $l=3,5,\dots,\frac{n}{2}-4$ 
which gives $\frac{1}{2}(\frac{n}{2}-5)$
different diagrams. 
In the seventh family the diagrams $D^{\text{sub},(1)}_{7,l}$
are the same as $D^{\text{sub},(1)}_{7,\frac{n}{2}-2-l}$
such that $l=3,5,\dots,\frac{n}{4}-2$ 
(or $l=3,5,\dots,\frac{n}{4}-1$) if $\frac{n}{4}$ is odd (even)
which gives $\frac{1}{2}(\frac{n}{4}-3)$ 
(or $\frac{1}{2}(\frac{n}{4}-1$)) different diagrams. If $\frac{n}{4}$ is
even and $l=\frac{n}{4}-1$ the multiplicity factor should be $\frac{n}{4}$
instead of $\frac{n}{2}$ as given in the formula above. In both cases
the sum of the multiplicity factors over all different diagrams
is $\frac{n}{4}(\frac{n}{4}-3)$ for the seventh family.

The non-complete diagrams of the sixth and seventh families are given by
\begin{equation}
  \begin{array}{rcl} 
    \hat{D}^{(1)}_{6a,l}=&
    \begin{array}{c}
      \includegraphics[scale=0.7]{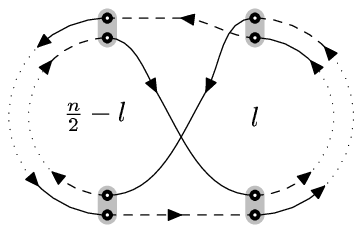} 
    \end{array}&
    =
    \frac{n}{2B} \hat{C}^{(1)}_{6a,l} 
    \\
    \hat{D}^{(1)}_{6b,l}=&
    \begin{array}{c}
      \includegraphics[scale=0.7]{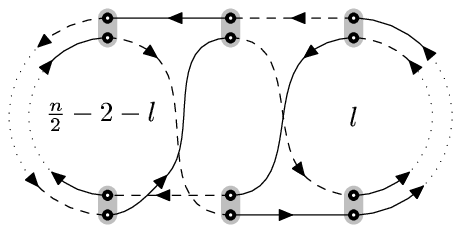} 
    \end{array}
    &
    =\frac{n}{2B} \hat{C}^{(1)}_{6b,l}
  \end{array}
  \label{eq:beyond_sd6a}
\end{equation}
and
\begin{equation}
  \begin{array}{rcl} 
    \hat{D}^{(1)}_{7a,l}=&
    \begin{array}{c}
      \includegraphics[scale=0.7]{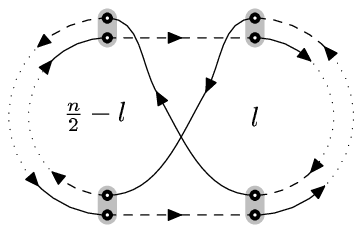} 
    \end{array}&
    =
    \frac{n}{2B} \hat{C}^{(1)}_{7a,l} 
    \\
    \hat{D}^{(1)}_{7b,l}=&
    \begin{array}{c}
      \includegraphics[scale=0.7]{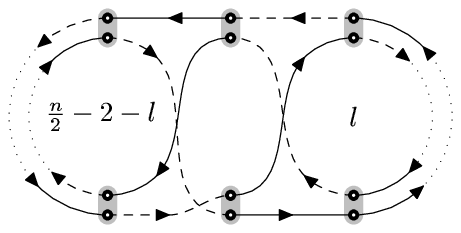} 
    \end{array}
    &
    =\frac{n}{2B} \hat{C}^{(1)}_{7b,l}.
  \end{array}
  \label{eq:beyond_sd7a}
\end{equation}
In both families $l$ is odd while $\frac{n}{2}$ is odd (even)
in the sixth (seventh) family.
The diagrams $D^{(1)}_{6a,1}$, $D^{(1)}_{6a,3}$,
$D^{(1)}_{6a,\frac{n}{2}-2}$, $D^{(1)}_{6b,1}$,
$D^{(1)}_{6b,\frac{n}{2}-2}$, $D^{(1)}_{7a,1}$, $D^{(1)}_{7a,3}$
and $D^{(1)}_{7b,1}$
are not new (they can be found among $D^{(1)}_{1a/1b,l}$ and
$D^{(1)}_{2a/2b,l,k}$).  
Thus $l=5,7,\dots,\frac{n}{2}-4$
(or $l=3,7,\dots,\frac{n}{2}-4$) for $D^{(1)}_{6a,l}$ ($D^{(1)}_{6b,l}$)
which gives $\frac{1}{2}(\frac{n}{2}-7)$  (or $\frac{1}{2}(\frac{n}{2}-5)$) 
new diagrams in the sixth family.

For odd (even) $\frac{n}{4}$ the multiplicity factor of the diagram
$D^{(1)}_{7a,\frac{n}{4}}$ ($D^{(1)}_{7b,\frac{n}{4}-1}$) for
is reduced to $\frac{n}{4}$ due to the symmetry of these diagrams.
If $\frac{n}{4}$ is odd (even) $l=5,7,\dots,\frac{n}{4}-2$ 
(or $l=3,5,\dots,\frac{n}{4}-3$) for the diagrams 
$D^{(1)}_{7a,l}$ ($D^{(1)}_{7b,l}$) and the sum over the multiplicities
is $\frac{n}{4}(\frac{n}{4}-4)$ (or $\frac{n}{4}(\frac{n}{4}-2)$).

Assuming that within each family the quasi-spin factors have all the same
value $C^{(1)}_{6,7}$
almost all diagrams in the sixth and seventh families are
canceled by sub-diagrams in the corresponding families.
The mechanism is similar to the third, fourth, and fifth family.
The main difference is that some
of the sub-diagrams in the sixth and seventh family also
appear as sub-diagrams in the first and second families of diagrams
which leads to additional overcounting. Each of the 
sub-diagrams actually appears three times as a sub-diagram -- e.g.
$D^{\text{sub},(1)}_{6,l}$ for $\frac{n}{2}-6 \ge l \ge 5$ 
is sub-diagram of $D^{(1)}_{6a,l}$, $D^{(1)}_{6a,l-2}$
and   $D^{(1)}_{6b,l}$ thus each has been overcounted twice.
The sum
over all diagrams with corresponding corrections due
to overcounting gives
\begin{subequations}
  \begin{align}
    \sum_{l} D^{(1)}_{6a,l}+\sum_l D^{(1)}_{6b,l} -2
    \sum_{l} D^{\text{sub},(1)}_{6,l}&=-\frac{n}{2B} C^{(1)}_{6}
    \label{eq:weaklocalization6}\\
    \sum_{l} D^{(1)}_{7a,l}+\sum_l D^{(1)}_{7b,l} -2
    \sum_{l} D^{\text{sub},(1)}_{7,l}&=-\frac{n}{4B} C^{(1)}_{7}\! ,\!
    \label{eq:weaklocalization7}
  \end{align}
\end{subequations} 
where the contribution of the sixth family only exists for odd $\frac{n}{2}$
and the contributions of the seventh family only for even $\frac{n}{2}$. 
These will be responsible for the leading order beyond
the self-dual approximation in all seven ensembles.

\subsubsection{The Andreev class $C$}

In class $C$ there are no additional spin-components of
the wave function. The ensemble
average leads to the condition that every scattering region
has as many incoming electron lines as hole lines and 
$C=\pm 2$ for all quasi-spin factors. The sign is positive for even 
$\frac{n}{2}$ and negative for odd $\frac{n}{2}$ 
(two d-vertices within a scattering region carry a factor $-1$) 
while the factor two 
corresponds to interchanging electron and hole lines. 
In the self-dual approximation only the diagram $D^{(0)}_1=-2$
fulfills the stated condition which leads to
\begin{equation}
  K_{1,n}=
  \begin{cases}
    0 & \text{$n=2s+1$}\\
    0 +\mathcal{O}(\frac{1}{B})& \text{$n=4s$}\\
    -4 +\mathcal{O}(\frac{1}{B})& \text{$n=2(2s+1)$}   
  \end{cases}
\end{equation}
where $s$ is some integer. Time-averaging gives the correct leading order
$K_1(\tau)=1+\mathcal{O}(\tau)$. In the next order all
diagrams in the first five families are canceled due to overcounting since
the quasi-spin factors are equal while in the sixth and seventh family
(and also some among the other five) each diagram contains at least
one loop with antiparallel lines. Along these any scattering region
has two incoming lines of the same type. thus they
do not survive the ensemble average and we have 
$K_1(\tau)=1+\mathcal{O}(\tau^2)$ as predicted by random-matrix theory 
\eqref{eq:novel_RMT}. 

\subsubsection{The Andreev class $C$I}

In class $C$I the quasi-spin factors are 
the same as for class $C$ as no spin is present. 
The ensemble average leads to the weaker condition that every
scattering region contains an even number of d-vertices.
In the self-dual approximation thus
both diagrams contribute. In the next to leading order only
the contribution from the sixth and seventh family contribute.
Altogether this gives
\begin{equation}
  K_{1,n}=
  \begin{cases}
    0 & \text{$n=2s+1$}\\
    2n-\frac{n}{B} 
    +\mathcal{O}(\frac{1}{B^2})& \text{$n=4s$}\\
    -4-2n+\frac{2n}{B} 
    +\mathcal{O}(\frac{1}{B^2})& \text{$n=2(2s+1)$}.   
  \end{cases}
\end{equation}
Since $\tau=\frac{n}{2B}$ time averaging yields
$K_1(\tau)=-1+\frac{\tau}{2} +\mathcal{O}(\tau^2)$
in accordance with the random-matrix theory result \eqref{eq:novel_RMT}..

\subsubsection{The Andreev class $D$}

In class $D$ we have to take the spin components into account. The ensemble
average lead to a set of conditions on the scattering
regions. It will suffice
to consider one large complete open loop to identify all
allowed configurations. The complete open
loops can be obtained from the two self-dual
diagrams by cutting two (anti-)parallel lines.
We will neglected contributions which are exponentially
suppressed in the limit $n,B \rightarrow \infty$. 
First consider an anti-parallel loop and put either
two parallel or
two anti-parallel spins on two anti-parallel lines.
For anti-parallel lines with parallel spins both spins are always flipped from
one side of the scattering region to the other. Anti-parallel spins on anti-parallel
lines are never flipped. In both cases we can only choose the spins
on one pair of parallel lines and the 
spins on \emph{all} lines in the loop are fixed. 
Since ever scattering region
carries a factor $\pm \frac{1}{2}$ any configuration with anti-parallel 
lines is suppressed exponentially with a factor $2^{-m}$ for
a loop of length $m$.

In a loop of parallel lines the situation is different.
Only configurations of parallel spins on parallel lines
survive the ensemble average. However the spins may either
flip or not when parallel lines hit a scattering region. 
Since there are $2^m$ such configurations on a loop
with $m$ scattering regions and $2m$ lines, the factor $2^{-m}$ of
the scattering regions  is canceled (in this configuration 
the factor from each scattering region is positive). For a loop we thus have
the same conditions on a scattering region as in class $C$. Only the diagrams
$D^{(0)}_1$, $D^{(1)}_{1b,l}$, $D^{(1)}_{3a,l}$
and $D^{(1)}_{3b,l}$ fulfill this condition. However,
the contributions of the first and third families have been shown
to vanish under the given conditions such that
contributions to the order $\tau$ remain. Altogether
\begin{equation}
  K_{1,n}=2\langle s_n \rangle=
  \begin{cases}
    0 & \text{$n=2s+1$}\\
    0 +\mathcal{O}(\frac{1}{B^2})& \text{$n=4s$}\\
    4 +\mathcal{O}(\frac{1}{B^2})& \text{$n=2(2s-1)$}   
  \end{cases}
\end{equation}
and time averaging yields the corresponding result from
random-matrix theory $K_1(\tau)=1+\mathcal{O}(\tau^2)$
upto the order we have calculated \eqref{eq:novel_RMT}.

\subsubsection{The Andreev class $D$III}

After the ensemble average in class $D$III the   
the spins have to be parallel in a parallel loop  
where the scattering regions carry a positive factor $\frac{1}{2}$.
In consequence, the first self-dual diagram has the value
$D^{(0)}_1=2$ as the quasi-spin factor is $C^{(0)}=2\, 2^\frac{n}{2}\,
2^{-\frac{n}{2}}=2$.  
For an anti-parallel loop
only configurations with anti-parallel spin need to be counted
(the sum over some remaining contributions is exponentially
suppressed). At any scattering region the spins may then either 
flip or not -- every time they both flip the scattering region carries a
negative sign (else the factor is positive). A turning point
inside a loop has anti-parallel spins on two connected lines
and always carries a positive sign whatever allowed
spin configuration on both sides. 
As a consequence, for any quasi-spin configuration that
contributes to $D^{(0)}_2$ there is another configuration with
opposite sign which has a different spin on one pair of 
lines connected to the  turning point region. We thus have  $D^{(0)}_2=0$.
The same is true for any diagram which contains a turning point.

The argument for the cancellation of
almost all diagrams but the contributions 
given in equations \eqref{eq:weaklocalization6} 
and \eqref{eq:weaklocalization7} 
in the preceeding section 
assumed the same quasi-spin factor for all diagrams.
It can
be generalized to class $D$III (and also all other classes)
if one properly only subtracts those quasi-spin configurations
in the sub-diagrams that actually have been overcounted (the
sub-diagrams may contain allowed configurations that have been counted
properly). The overcounted configurations that appear in
the quasi-spin factors $C^{(1)}_{6,7}$ are the corresponding quasi-spin factors 
of the diagrams $D^{(1)}_{6a,l}$ and $D^{(1)}_{7a,l}$. For configurations
that contribute to $C^{(1)}_{6}$ the number of spin flips in the left loop
is necessarily positive which gives a positive sign to each contribution.
Altogether one gets $C^{(1)}_{6}=1$. Indeed,
all the scattering regions give a factor $2^{-\frac{n}{2}}$, the
electron-hole interchange gives a factor $2$ and the spins $2^{\frac{n}{2}-1}$
(the spins on the four lines connecting the two loops are determined
by a single spin index). 
The other quasi-spin factor is $C^{(1)}_{7}=-1$. The different sign
is due to an odd number of spin flips along both anti-parallel loops
here.
The complete result for the form factor $K_{1,n}=\langle s_n \rangle$ is
\begin{equation}
  K_{1,n}=
  \begin{cases}
    0 & \text{$n=2s+1$}\\
    -\frac{n}{2B}+
    \mathcal{O}(\frac{1}{B^2})& \text{$n=4s$}\\
    2+\frac{n}{4B}
    \mathcal{O}(\frac{1}{B^2})& \text{$n=2(2s+1)$}   
  \end{cases}.
\end{equation}
As $\tau=\frac{n}{2B}$ time averaging yields
$K_1(\tau)=\frac{1}{2}-\frac{\tau}{8}
  +\mathcal{O}(\tau^2)$
which is again the corresponding random-matrix theory result \eqref{eq:novel_RMT}.

\subsubsection{The chiral class $A$III}

Parallel loops do not
survive  the ensemble average in the chiral class $A$III. Thus the first 
self-dual diagram vanishes $D^{(0)}_1=0$. Long loops
with anti-parallel lines only have weight if the
spins are always parallel. At each scattering region
they may either flip or
not -- in both cases the scattering region carries a positive factor.
Turning points inside a loop carry a negative sign
if the two incoming lines have opposite spin, else the sign is positive.
By flipping spins on one side of a turning point one can thus change
the overall sign of a configuration. Eventually all diagrams with a turning
point vanish. There is no contribution at all to the
self-dual approximation. Among all diagrams that contribute to
the linear order only the seventh family has to be considered --
all other diagrams either contain a turning point or a 
long loop of parallel lines or they cancel due to overcounting.
In equation \eqref{eq:weaklocalization7} the factor $C^{(1)}_7$
is the quasi-spin factor of the diagrams $D^{(1)}_{7,l}$
which is $C^{(1)}_7=1$. Altogether we
have
\begin{equation}
  K_{1,n}=
  \begin{cases}
    0 & \text{$n=2s+1$}\\
    0+
    \mathcal{O}(\frac{1}{B^2})& \text{$n=4s$}\\
    -\frac{n}{2B}+
    \mathcal{O}(\frac{1}{B^2})& \text{$n=2(2s+1)$}.   
  \end{cases}
\end{equation}
Here $\tau=\frac{n}{4B}$ such that time averaging gives
the form factor $K_1(\tau)=-\frac{\tau}{2}
  +\mathcal{O}(\tau^2)$
as predicted by random-matrix theory \eqref{eq:novel_RMT}.

\subsubsection{The chiral class $BD$I}

In the next chiral class, $BD$I  long loops of parallel lines
survive the ensemble average
in addition to the anti-parallel loops of class $A$III.
For loops of parallel lines all scattering regions carry a positive sign.
For anti-parallel loops and turning points the discussion of class $A$III
can be taken over completely. The only additional
contributions to the form factor are due to the first self-dual
diagram $D^{(0)}_1=2$ and due to the contribution
\eqref{eq:weaklocalization6} of the sixth family of diagrams for
the linear order. Here, $C^{(1)}_6=1$ and we arrive at 
\begin{equation}
  K_{1,n}=
  \begin{cases}
    0 & \text{$n=2s+1$}\\
    0-\frac{n}{2B}+
    \mathcal{O}(\frac{1}{B^2})& \text{$n=4s$}\\
    4-\frac{n}{B}+
    \mathcal{O}(\frac{1}{B^2})& \text{$n=2(2s+1)$}   
  \end{cases}.
\end{equation}
Again, with $\tau=\frac{n}{4B}$ and  time averaging 
we get the random-matrix result \eqref{eq:novel_RMT}
$K_1(\tau)=1-\frac{3\tau}{2}
  +\mathcal{O}(\tau^2)$.

\subsubsection{The chiral class $C$II}

Finally, in class $C$II the discussion is almost
equivalent to the preceeding. Loops of parallel lines have anti-parallel
spins and in such a loop a 
scattering region carries a negative sign if both spins flip.
In conclusion $D^{(0)}_1=-2$ since an odd number of spin flips occurs.
For anti-parallel lines the spins are always parallel and
the scattering regions carry a positive sign. Turning points
inside such a loop can have either sign.
For $C^{(1)}_6$ the number of spin flips 
along the parallel loop is always even, thus  $C^{(1)}_6=C^{(1)}_7=1$
and we have
\begin{equation}
  K_{1,n}=
  \begin{cases}
    0 & \text{$n=2s+1$}\\
    0-\frac{n}{4B}+
    \mathcal{O}(\frac{1}{B^2})& \text{$n=4s$}\\
    -2-\frac{n}{2B}+
    \mathcal{O}(\frac{1}{B^2})& \text{$n=2(2s+1)$}.   
  \end{cases}
\end{equation}
Here, $\tau=\frac{n}{2B}$ and the time average yields 
$K_1(\tau)=-\frac{1}{2}-\frac{3\tau}{8}
  +\mathcal{O}(\tau^2)$.
Needles to say this is in accordance with the random-matrix theory
prediction \eqref{eq:novel_RMT}.

\section{Conclusion}

We have given a systematic diagrammatic short-time expansion of the
first-order and second-order form factors for ensembles of
star graphs in the ten symmetry classes. The leading orders
(diagonal and self-dual approximations) have been calculated explicitly
along with the first order beyond. The fidelity to the
predictions of Gaussian random-matrix ensemble has been established
to this order. These results support the proper generalization
of the Bohigas-Giannoni-Schmit conjecture to the novel symmetry classes.
The contributing diagrams for the ensembles of star graphs can be expected
to carry over to Hamiltonian flows. For magnetic Andreev billiards in class $C$
this is indeed the case \cite{us}. For more general flows in all
novel classes a description of
the self-dual approximation will be given in a future work \cite{flows}. 
A theory for
flows beyond the self-dual approximation for the novel ensembles
will follow the paths of the
Sieber-Richter theory for 
time-reversal invariant Wigner-Dyson systems. 
The results given here for quantum graphs are
however more systematic as the existing work on Wigner-Dyson flows as
we could show that no other contributions exist that contribute to
the calculated order of the form factors.
 
\acknowledgments

We are indebted to Felix von Oppen and Martin Zirnbauer
for many helpful suggestions, comments and discussions. 
We thank for the support
of the Sonderforschungsbereisch/Transregio 12 of the Deutsche 
Forschungsgemeinschaft.

\end{document}